\documentclass[fleqn,usenatbib]{mnras}

\usepackage{newtxtext,newtxmath}
\usepackage[T1]{fontenc}

\DeclareRobustCommand{\VAN}[3]{#2}
\let\VANthebibliography\thebibliography
\def\thebibliography{\DeclareRobustCommand{\VAN}[3]{##3}\VANthebibliography}

\usepackage{graphicx}	% Including figure files
\usepackage{amsmath}	% Advanced maths commands
\usepackage{xcolor}
\usepackage{xspace}
\newcommand{\python}{\textsc{python}\xspace}
\newcommand{\afterglowpy}{\textsc{afterglowpy}\xspace}
\newcommand{\dynesty}{\textsc{dynesty}\xspace}
\newcommand{\bilby}{\textsc{bilby}\xspace}
\newcommand{\xspec}{\textsc{xspec}\xspace}

\usepackage{gensymb}
\title[Multimessenger study of GW170817]{Joint  analysis of gravitational-wave and electromagnetic data of mergers: breaking an afterglow model degeneracy in GW170817 and in future events}

\author[G. Gianfagna et al.]{%
Giulia Gianfagna,$^{1,2}$\thanks{E-mail: giulia.gianfagna@inaf.it}
Luigi Piro,$^{1}$
Francesco Pannarale,$^{2,3}$
Hendrik Van Eerten,$^{4}$
Fulvio Ricci,$^{2,3}$
\newauthor
Geoffrey Ryan,$^{5}$
Eleonora Troja$^{6,7}$\\
$^{1}$INAF - Istituto di Astrofisica e Planetologia Spaziali, via Fosso del Cavaliere 100, 00133 Rome, Italy\\
$^{2}$Dipartimento di Fisica, Università di Roma "Sapienza", Piazzale A. Moro 5, I-00185, Roma, Italy\\
$^{3}$INFN Sezione di Roma, Piazzale A. Moro 5, I-00185, Roma, Italy\\
$^{4}$Department of Physics, University of Bath, Claverton Down, Bath, BA2 7AY, UK\\
$^{5}$Perimeter Institute for Theoretical Physics, 31 Caroline St. N., Waterloo, ON, N2L 2Y5, Canada\\
$^{6}$Astrophysics Science Division, NASA Goddard Space Flight Center, 8800 Greenbelt Rd, Greenbelt, MD 20771, USA\\
$^{7}$University of Rome Tor Vergata, Department of Physics, via della Ricerca Scientifica 1, 00100, Rome, IT
}

\date{Accepted XXX. Received YYY; in original form ZZZ}

\pubyear{2022}

\begin{document}
\label{firstpage}
\pagerange{\pageref{firstpage}--\pageref{lastpage}}
\maketitle

\begin{abstract}
On August 17, 2017, Advanced LIGO and Virgo observed GW170817, the first gravitational-wave (GW) signal from a binary neutron star merger. It was followed by a short-duration gamma-ray burst, GRB 170817A, and by a non-thermal afterglow emission.
In this work, a combined simultaneous fit of the electromagnetic (EM, specifically, afterglow) and GW domains is implemented, both using the posterior distribution of a GW standalone analysis as prior distribution to separately process the EM data, and fitting the EM and GW domains simultaneously. These approaches coincide mathematically, as long as the actual posterior of the GW analysis, and not an approximation, is used as prior for the EM analysis.
 We treat the viewing angle, $\theta_v$, as shared parameter across the two domains.
In the afterglow modelling with a Gaussian structured jet this parameter and the jet core angle, $\theta_c$, are correlated, leading to high uncertainties on their values. 
The joint EM+GW analysis relaxes this degeneracy, reducing the uncertainty compared to an EM-only fit.
We also apply our methodology to hypothetical GW170817-like events occurring in the next GW observing run at $\sim$140 and 70\,Mpc. 
At 70\,Mpc the existing EM degeneracy is broken, thanks to the inclusion of the GW domain in the analysis.
%placing strong constraints on both $\theta_v$ and $\theta_c$.
At 140\,Mpc, the EM-only fit cannot constrain $\theta_v$ nor $\theta_c$ because of the lack of detections in the afterglow rising phase. Folding the GW data into the analysis leads to tighter constraints on $\theta_v$, still leaving $\theta_c$ unconstrained, requiring instruments with higher sensitivities, such as \textit{Athena}.
%250 words

\end{abstract}


%%%%%%%%%%%%%%%%%%%%%%%%%%%%%%%%%%%%%%%%%%%%%%%%%%

%%%%%%%%%%%%%%%%% BODY OF PAPER %%%%%%%%%%%%%%%%%%

\section{Introduction}
\label{sec:intro}

On August 17, 2017, the two Advanced LIGO detectors \citep{AdvancedLIGO2015} and Advanced Virgo \citep{AdvancedVIRGO2015}
observed the first neutron star binary inspiral event GW170817 \citep{Abbott2017_1}. The collisions of two neutron stars form highly-relativistic and collimated outflows (jets) that power gamma-ray bursts (GRBs) of short ($\lesssim 2\,$s) duration \citep{Blinnikov1984, Paczynski1986, Eichler1989, Paczynski:1991aq, Narayan1992, nakar2007, Berger2014}. Therefore, gravitational-wave (GW) events from such mergers should be associated with GRBs, but the majority of these bursts should be viewed off-axis, that is, they should point away from Earth \citep{Rhoads1997, Piran2005, Nakar2020}. 

The joint detection of the event GW170817 from a binary neutron star merger \citep{Abbott2017_1} and of the short, hard burst GRB 170817A \citep{Goldstein2017, Savchenko2017} confirmed the compact binary progenitor model for at least some short GRBs \citep{Abbott2017_dis, Ashton:2017ykh}. Observations in the X-ray \citep{Troja2017_Nature} and, later, radio frequencies \citep{Hallinan2017} are consistent with a short GRB viewed off-axis \citep{vanEerten2010}. The general findings are that a relativistic jet was launched, that it successfully broke out of the ejecta, and that it interacted with the circumburst medium to produce the afterglow. The jet had a narrow core, and we observed it with a viewing angle larger than its opening angle \citep[e.g.,][]{Troja2017_Nature, Margutti2017, Haggard2017, Alexander2018, gill2018, Dobie2018, Granot2018, DAvanzo2018, Lazzati2018, Lyman2018, Margutti2018, Mooley2018, Troja2018, Fong2019, Hajela2019, Lamb2019, Wu2019, Ryan2020, Troja2020,  troja2021, Takahashi2021, Hajela2022}. 

Following the detection of GW170817, several multi-messenger studies combined GW and electromagnetic (EM) data. This enables us to refine our understanding of the physics of binary neutron star merger events and to break degeneracies encountered when processing the EM and GW data independently.
One vein followed by such studies pertains to the neutron star equation of state.
Bayesian methods are implemented to combine GW170817 and the associated kilonova AT2017gfo, in order to provide improved estimates of source parameters such as the mass ratio and the neutron star tidal deformability \citep[e.g.,][]{Radice2018, Hinderer2019, Radice2019, Capano2020, Raaijmakers2021, Raaijmakers2021b, Coughlin2018, Coughlin2019, Dietrich2020, Breschi2021, Nicholl2021}. This is done by exploiting the fact that, for a wide range of neutron star equation of state models, the dynamical ejecta mass is essentially determined by the spin and the mass ratio of the two neutron stars \citep[e.g.,][]{Hotokezaka2013, Sekiguchi2016}, which can be constrained along with the neutron star tidal deformability with GW data from the inspiral stage of neutron star mergers \citep[e.g.,][]{Read2013}, while kilonova emission models in the EM domain place constraints on the ejecta mass \citep[e.g.,][]{Bernuzzi2015, Villar2017, Kasen2017}. 

A second vein, that is at the core of this paper, focuses on the system geometry, which includes the viewing angle and the angular structure of the jet. There have been numerous attempts to estimate the system geometry of GW170817 based on model fitting of the EM afterglow light curve, since the early discovery papers of the afterglow in the X-ray and radio bands \citep[and references therein]{Troja2017_Nature, Hallinan2017}. The viewing angles and jet opening angles found in different studies are often inconsistent with each other. 
 
As pointed out by \citet{Ryan2020}, assuming a Gaussian structure of the jet, the rising light curve phase can only constrain the ratio between viewing angle and jet opening angle. This leads to an intrinsic degeneracy between these parameters: a proper scaling of the two angles leads to the same light curve, with some slight changes in the decreasing slope. Breaking this degeneracy requires further information (see also \citealt{Nakar2021} for additional discussion).

One way of improving the precision of the viewing angle measurement is to break the viewing angle--jet opening angle degeneracy. 
Within the framework of the same messenger, one option is to join the GW170817 afterglow measurements with independent measurements, such as the observations that measure the superluminal motion of the centroid in radio and optical images \citep{Mooley2018, Ghirlanda2019, Mooley2022}, or the late time observations of the light curve transition to the sub-relativistic phase \citep{Ryan2020}. However, these measurements can be available only for very close events, like GW170817.
A second option is to join information from GW and EM counterparts, either by jointly fitting the two datasets or by including GW information in the EM analysis as a constraint on the common parameters, either a priori or a posteriori.  
\citet{Wang2021} use a Bayesian approach to fit the observed GW170817 afterglow emission with a simulated jet profile that predicts the afterglow light curve.  This is done to avoid the degeneracy of the observing angle with the jet opening angle in prescribed jet structure models (e.g., Gaussian, power-law, etc.). They impose a GW-informed prior on the luminosity distance and the observing angle. The latter can be tightly constrained and the distance uncertainty is in turn also greatly reduced. This results in a tight constraint on the observing angle $\theta_v = 22\degree \pm 1 \degree$.
\citet{hotokezaka2018} performed the modelling of the GW170817 afterglow light curve and centroid motion data, taking into account the constraint from GW data. They use the posterior distribution of the viewing angle and luminosity distance from the GW analysis as a prior for the EM modelling, finding a viewing angle of 15--29$\degree$. 
\citet{Troja2018, Troja2019, Troja2020, troja2021} study the broadband afterglow of GW170817 with a Bayesian fit, imposing a GW-informed prior \citep{Abbott2017_dl} on the viewing angle and fixing the luminosity distance. The viewing angle goes from almost 20$\degree$ (2018) to almost 30$\degree$ (2021), because of the relative excess of the late-time X-ray observations.
\citet{Guidorzi2017} use a Bayesian analysis and reanalyze the LIGO-Virgo posteriors incorporating the viewing angle constraints obtained from the X-ray and radio afterglow modeling, finding an off-axis angle of 25--50$\degree$.

In this paper we implement, for the first time, a truly joint Bayesian fit of the GW and EM data for the event GW170817. The model used to process the data involves gravitational waveforms emitted by binary coalescences for the GW signal \citep{bilby_paper, pbilby_paper} and a Gaussian structured jet model \citep{Ryan2020} for the EM data.
We test our approach on GW170817 data and on EM data from its afterglow counterpart, and compare our results with those derived adopting the approach of using disjoint priors in the two domains.  We focus on the geometry of the system, and specifically on the issue on the degeneracy between the viewing angle and the opening angle of the jet. Our goal is to compare the EM fit with a GW-informed prior and an EM+GW fit, showing also that the methods implemented here provide a significant improvement over disjoint analyses. 
In this context, we keep the luminosity distance fixed, compliant with the procedure usually adopted in the analysis of  afterglow properties.  
We note that, for the purpose of assessing $H_0$, the luminosity distance is left as free parameter, as implemented for example in \citet[]{Fan2014, Guidorzi2017, hotokezaka2018, Finstad2018, Dietrich2020, Biscoveanu2020, Wang2021, Wang2023}, see also the recent review by \citet[]{Bulla2022}.

In Section \ref{sec:methods} we introduce our method, and how we combine the two datasets. In Section \ref{sec:results} we present our results, discussing future perspectives for O4, the upcoming GW observing run, in Section \ref{sec:results_O4}. Finally, in Section \ref{sec:conclusions} we provide our conclusions. Throughout our analysis of GW170817 we adopt the redshift value $z=0.0098$ and set the luminosity distance to $d_{\rm L} = 41$ Mpc \citep{Abbott2017_dl}.

\section{Methodology}
\label{sec:methods}

This section details the methods and data used in this work. We choose to fix the cosmology (i.e., to fix the luminosity distance) in the EM fits to analyse the data in the EM standard fashion. For consistency, we fix the cosmology also for the GW fits: this allows us to safely compare results from EM and GW fits, as well as from a simultaneous fit of the EM \emph{and} GW data.  

\subsection{Data}

\begin{figure}
    \includegraphics[width=0.5\textwidth]{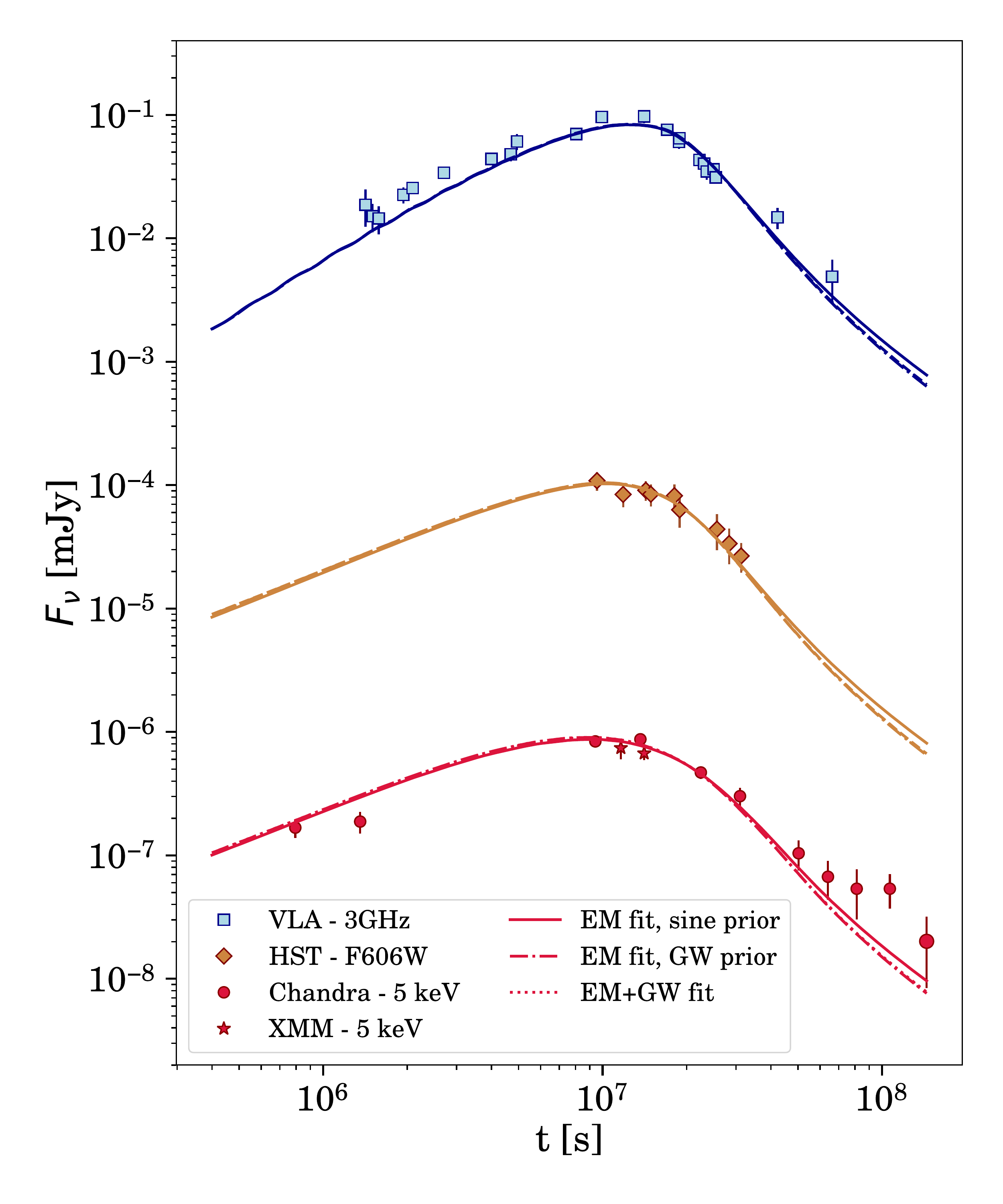}
    \caption{Broad-band afterglow of GW170817: data and fits. From bottom to top, red points refer to the X-ray observations by \textit{Chandra} and \textit{XMM} at 5\,keV, orange diamonds to observations by \textit{HST}, F606W filter, in the optical band, and blue squares to observations in the radio band from VLA at 3\,GHz. The lines represent fits of the afterglow data: continuous lines are obtained with a prior uniform in the cosine of $\theta_{v}$ (sine prior), dot-dashed lines with a GW-informed prior, and dotted lines with a simultaneous fit of EM and GW data. The last two models are indistinguishable. For sake of simplicity, the fit for the radio band is plotted only for the observations at 3\,GHz, but it is not limited to this single frequency.}
     \label{Fig:lightcurve_gw170817}
\end{figure}

We use the afterglow X-ray emission of GW170817 from \textit{Chandra} and \textit{XMM}, already introduced in \citet{troja2021}, and represented in Fig.\,\ref{Fig:lightcurve_gw170817} with red circles. The most recent Chandra data point at 1734 days from the merger is also included in the analysis \citep[][]{GCN_Chandra}. For the optical light curve, we refer to \citet{Troja2017_Nature} and \citet{Fong2019}, see the orange diamonds in Fig.\,\ref{Fig:lightcurve_gw170817}. Finally, for the radio dataset we refer to \citet{makhathini2021}; in Fig.\,\ref{Fig:lightcurve_gw170817} we show only the VLA detections at 3\,GHz for the sake of good order, but include in the fit all frequencies from 0.7 to 15 GHz for VLA, ATCA, uGMRT, eMERLIN, MeerKAT. 

The GW data of GW170817 are publicly available at the GW Open Science Center\footnote{\href{https://www.gw-openscience.org/}{www.gw-openscience.org}}\citep{LIGOScientific:2019lzm}. We use the cleaned version of the strain data, where the glitch discussed in \citet{Abbott2017_1} has been removed. 
%the psd data are from ...

\subsection{Afterglow data analysis}
\label{sec:methods-afterglow}

The afterglow light curve of GRB 170817A is modelled using \afterglowpy \citep{Ryan2020}.  
This \python package can compute the observer frame flux of synchrotron radiation for various jet geometries, including the Gaussian distribution of lateral energy used in this paper. For the Gaussian structured jet, energy drops according to $E(\theta) = E_0 \exp(-\theta^2 / 2\theta_c^2)$, up to a truncating angle $\theta_w$. $E_0$, $\theta_c$ and $\theta_w$ are treated as free parameters and they represent the on-axis isotropic equivalent kinetic energy of the blast wave, the jet opening angle and the jet total angular width, respectively. We assume that the electrons are shock-accelerated and emit synchrotron radiation, their energy distribution is a power law with slope $-p$, the fraction of their post-shock internal energy is $\epsilon_e$, while the fraction of post-shock internal energy in the magnetic field is denoted by $\epsilon_B$. Furthermore, the circumburst medium number density $n_0$ and the viewing angle $\theta_v$, between the jet axis and the line of sight, are free parameters as well. This yields a total of 8 free parameters, given that the luminosity distance and the redshift are fixed (see Section \ref{sec:intro}). The participation fraction $\chi_{N}$ is fixed to 1.0 as well.

The prior probability distributions are uniform for $\theta_w$ in [0$\degree$, 90$\degree$], $\theta_c$ in [0$\degree$, 90$\degree$], and $p$ in [2, 3]; log-uniform for $E_0$ in [49, 56], $\epsilon_e$ in [-5,0], $\epsilon_B$ in [-5,0], and $n_0$ in [-7,2]. We use two prior distributions for the viewing angle $\theta_v$: an isotropic one, which has a sinusoidal shape from 0$\degree$ to 90$\degree$, and a GW-informed one, which is the posterior distribution retrieved from fitting the GW data alone (see Section\,\ref{sec:methods-GW}).

\subsection{Gravitational-wave data analysis}
\label{sec:methods-GW}

We use the Bayesian inference library for GW astronomy \bilby \citep{bilby_paper, pbilby_paper} and the dynamic nested sampling package \dynesty \citep{Speagle2020} to process the GW data. To model GWs from binary neutron star mergers, we use the {\ttfamily IMRPhenomPv2\_NRTidal} waveform model \citep{Hannam2014, Dietrich2017, Dietrich2019_2, Dietrich2019}. The parameters used by this model are the two component masses $m_1$ and $m_2$, for which we follow the common convention $m_1 \geq m_2$, the components of the dimensionless spin angular momenta of the two neutron stars, $\mathbf{a}_1$ and $\mathbf{a}_2$, which constitute six additional parameters, and the tidal deformability parameter of each star, $\Lambda_1$ and $\Lambda_2$, for a total of 10 parameters.  The tidal deformability parameter $\Lambda$ quantifies how an extended object responds to the $\ell=2$ component of an external tidal field; it is related to the mass $m$ and radius $R$ of the object by $\Lambda = (2/3) k_2 [(c^2/G) (R/m)]^5$, where $k_2$ is the dimensionless $\ell = 2$ Love number \citep[e.g.,][]{Yagi2014, Wade2014}.

In addition to the gravitational waveform physical parameters listed above, known as intrinsic parameters as they shape the \emph{emitted} waveform, the \emph{observed} GW signal is determined also by seven extrinsic parameters.  These are the right ascension and declination of the source (i.e., its sky position), the luminosity distance of the source, the inclination angle $\theta_{\rm JN}$ between the total angular momentum $\textbf{J}$ of the binary and the line of sight $\textbf{N}$ from the source to the observer (i.e., $\cos \theta_{\rm JN} = \hat{J} \cdot \hat{N}$), the polarization angle, and the phase and time of coalescence.  The total number of parameters is therefore 17.

In this work, we fix the sky-position of the source to the one of AT 2017gfo \citep{Abbott2017_position} both when processing GW170817 and when considering future GW170817-like events.  In other words, we address scenarios in which the EM observations have pinpointed the source.  Similarly, we also assume the luminosity distance to be known.  In the case of GW170817, we set this to $41\,$Mpc, while for the simulated future events we set it to $136.5\,$Mpc, which corresponds to a reduction of the afterglow flux of about one order of magnitude, and to 70\,Mpc.
Finally, we marginalize over the phase of coalescences, thus reducing the number of free GW parameters to 13. The priors for the intrinsic and extrinsic GW parameters are uniform for the chirp mass $\mathcal{M}$ in [1.18, 1.21], the mass ratio $q$ in [0.125, 1], the dimensionless spin magnitudes $a_1$ and $a_2$ in [0, 0.05], the opening angle of the cone of precession about the system angular momentum $\phi_{JL}$ and the azimuthal angle separating the spin vectors $\phi_{1,2}$ in [0$\degree$, 360$\degree$] with periodic boundary, the tidal deformability parameter $\Lambda_1$ and $\Lambda_2$ in [0, 5000]; sinusoidal for the tilt angles between the spins and the orbital angular momentum $\theta_1$, $\theta_2$ in [0$\degree$, 180$\degree$] \citep{Romero-shaw2020}.

\subsection{Joint gravitational-wave and afterglow data analysis}
\label{sec:methods-joint}

In the joint analysis there is a total of 21 parameters. However, given that the GRB jet develops around $\mathbf{J}$, the inclination angle $\theta_{\rm JN}$ introduced in Sec.\,\ref{sec:methods-GW} and the viewing angle $\theta_v$ introduced in Sec.\,\ref{sec:methods-afterglow} are essentially the same quantity, and thus a common parameter of the GW and EM domains. When $\textbf{J}$ points towards the observer, $\theta_{\rm JN}<90\degree$ and $\theta_v = \theta_{\rm JN}$, while when $\textbf{J}$ points away from the observer, $\theta_{\rm JN}>90\degree$ and $\theta_v = 180\degree - \theta_{\rm JN}$.  In summary,
\begin{equation}
\label{eq:theta_v_JN}
\theta_v = 90\degree -|\theta_{\rm JN} - 90\degree|\,.
\end{equation}
We denote the entire set of parameters with $\vec{\vartheta}$.

Assuming that the GW and EM datasets $d_{GW}$ and $d_{EM}$ are independent, their joint likelihood \citep[see also][]{Fan2014, Biscoveanu2020} is given by the product of the two likelihoods
\begin{equation}
    \mathcal{L}_{\textrm{EM+GW}}(d_{\textrm{EM}}, d_{\textrm{GW}}| \vec{\vartheta}) = \mathcal{L}_{\textrm{EM}}(d_{\textrm{EM}}| \vec{\vartheta}) \times \mathcal{L}_{\textrm{GW}}(d_{\textrm{GW}}| \vec{\vartheta})\,,
\label{Eq:likelihood}
\end{equation}
where, for notation simplicity and because we will not carry out hypotheses tests, we omit to explicitly indicate the assumptions about the gravitational waveform model and jet energy profile model discussed in the previous sections.
The EM and GW likelihoods are both Normal distributions. The GW likelihood function is defined in, e.g., \citet{Finn1992, Romano2017, Romero-shaw2020}; in this likelihood, both the data and the model are expressed in the frequency domain. The EM likelihood function is proportional to $\mathcal{L}_{\textrm{EM}}(d_{\textrm{EM}}| \vec{\vartheta}) \propto \exp (-{\chi }^{2}/2)$, where $\chi^2$ is evaluated from comparing the flux estimated from the model with $\vec{\vartheta}$ with the entire broadband set of data $d_{\rm EM}$.

Finally, the multi-dimensional posterior probability distribution for $\vec{\vartheta}$ is determined according to the Bayes theorem:
\begin{equation}
   p(\vec{\vartheta} |d_{\textrm{EM}}, d_{\textrm{GW}} ) \equiv \frac{\mathcal{L}_{\textrm{EM+GW}}(d_{\textrm{EM}}, d_{\textrm{GW}}| \vec{\vartheta}) \pi(\vec{\vartheta})}{\mathcal{Z}_{\vec{\vartheta}}}
\label{Eq:bayes_th_joint}
\end{equation}
where $\pi(\vec{\vartheta})$ is the multi-dimensional prior probability distribution for our 21 parameters. $\mathcal{Z}_{\vec{\vartheta}}$ is the Bayesian evidence, obtained by marginalizing the joint likelihood over the GRB and GW parameters
\begin{equation}
    \mathcal{Z}_{\vec{\vartheta}} = \int \mathcal{L}_{\textrm{EM+GW}}(d_{\textrm{EM}}, d_{\textrm{GW}}| \vec{\vartheta}) \pi(\vec{\vartheta}) d\vec{\vartheta}\,.
\end{equation}

In this work, we both fit the EM data using a GW-informed prior on the viewing angle, and we fit the EM and GW datasets together. These two fits are mathematically the same, see Appendix \ref{Sec:Appendix} for a detailed explanation. However, differences in the posteriors can arise if, in the case of the two separate fits (EM fit with a $\theta_v$ GW-informed prior), the prior on the common parameter in the EM fit does not accurately approximate the $\theta_{\rm JN}$ posterior of the GW fit, statistical noise deriving from the sampling aside. 

\begin{figure}
    \includegraphics[width=0.5\textwidth]{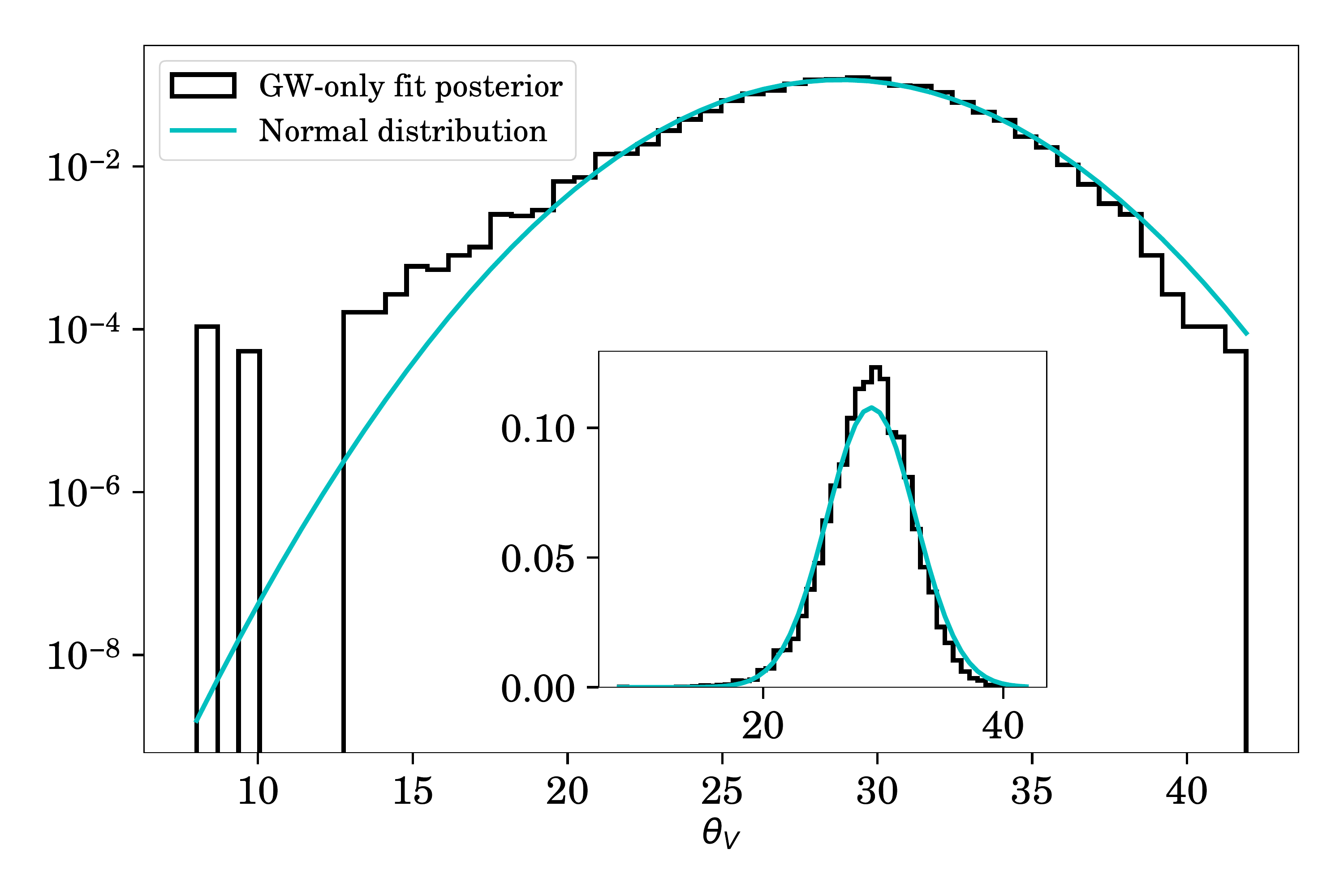}
    \caption{Comparison between the posterior distribution of the viewing angle for the GW-only fit (in black) and the Normal distribution with mean and standard deviation from the GW-posterior distribution ($\mu = 29.0\degree$ and $\sigma = 3.5\degree$), in cyan. In the subplot the same distributions are represented, but with a linear vertical axis.}
     \label{Fig:priors}
\end{figure}

\begin{figure}
    \includegraphics[width=0.48\textwidth]{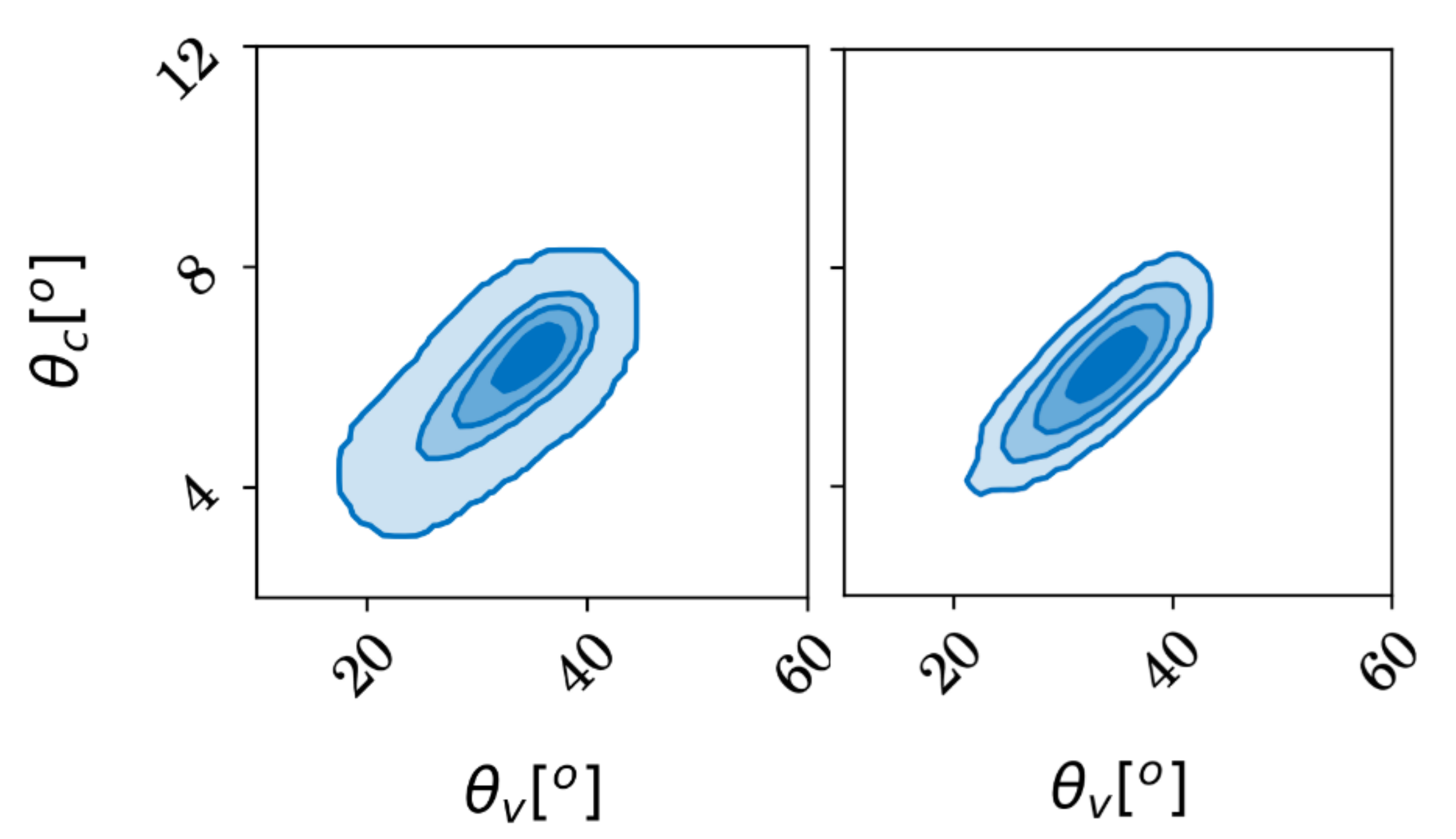}
    \caption{Posterior distributions of the jet opening angle $\theta_c$ and the viewing angle $\theta_v$ for an EM fit with a GW Normal prior (on the left) and the joint fit (on the right). The contour lines represent the 68.2\%, 95.4\%, 99.8\% and 99.98\% probabilities.}
     \label{Fig:posteriors_comparison}
\end{figure}

In Fig.\,\ref{Fig:priors} we plot two possible priors for the viewing angle: a Normal distribution with $\mu = 29.0\degree$ and $\sigma = 3.5\degree$ (estimated from the GW posterior distribution) and the $\theta_{\rm JN}$ posterior from the GW fit, transformed into $\theta_v$ according to Eq.\,(\ref{eq:theta_v_JN}). The main differences lay in the tails of the distributions and propagate in the posteriors. We show the results of our analysis, in particular the posteriors of $\theta_c$ and $\theta_v$ (see Section\,\ref{sec:results}), carried out using the Normal distribution as GW-informed prior on $\theta_v$, in Fig.\,\ref{Fig:posteriors_comparison}, left panel. They are compared with the results of the joint fit of the EM and GW datasets, in the right panel. While until 99.8\% probability the contours are similar, at 99.98\% there is a clear difference.

To quantify these differences, we refer to Table\,\ref{table:percentiles}. The first column represents the percentiles of each distribution. In the second column the differences between the $\theta_v$ Normal distribution and the GW posterior are written in percentages and refer to the GW posterior distribution, meaning that, for example, when the percentage is negative, the Normal distribution at that percentile predicts a larger angle with respect to the GW posterior distribution. In the third column we compare the results of the EM fit with the Normal GW-informed prior and the posterior from the EM+GW fit, in this case the percentages refer to the EM+GW fit. In the fourth column the results of the EM fit with the GW posterior as GW-informed prior and the posterior from the EM+GW fit are compared, also in this case the percentages refer to the EM+GW fit. This combination is the one used in Section\,\ref{sec:results} and \ref{sec:results_O4}. This last column, following the mathematical reasoning, should be 0, but it is not because of the statistical noise deriving from the sampling.

\begin{table}
    \caption{In this Table, in the second column, we report the differences between the $\theta_v$ Gaussian prior used in the EM fit with GW-informed prior and the posterior distribution on $\theta_v$ of the GW fit, \textit{(GW posterior - Normal distribution)/GW posterior}; in the third column the difference between the $\theta_v$ posterior distribution from the EM fit with GW-informed prior and the $\theta_v$ posterior of the EM+GW joint fit, \textit{(EM+GW posterior - EM posterior)/EM+GW posterior}; in the fourth column the differences between the posterior from the EM fit with the GW posterior as prior on the inclination and the $\theta_v$ posterior of the EM+GW fit, \textit{(EM+GW posterior - EM posterior)/EM+GW posterior}. The latter configuration is used in Sections \ref{sec:results} and \ref{sec:results_O4}. These quantities are represented as function of the percentiles of each distribution (first column).}
    \begin{tabular}{cccc} 
    \hline
    Percentile & Difference in Prior & Difference in Posterior & Main Paper \\ 
    
        [\%] & [\%] & [\%] & [\%] \\
        \hline
        0.01 & -7.9 & 9.2 & 4.2 \\
        0.1  & -4.2  & -6.4 & -4.0 \\
        2.3  & 3.0  & -2.0 & 1.2 \\
        15.9 & 3.1  & -2.4 & 1.6 \\
        50   & 1.0  & -2.4 & 2.1 \\
        84.1 & -1.5 & -1.0 & 2.0 \\
        97.7 & -3.4 & 2.0  & 1.5 \\
        99.9 & -5.3 & 6.4  & 3.0 \\
        99.99& -4.1 & 6.1  & 2.1 \\
    \hline
\end{tabular}
\label{table:percentiles}
\end{table}

From this Table, if we focus on the prior comparison, we can see that the differences between the Normal distribution and the actual GW posterior for $\theta_{\rm JN}$ do not go over 8\%. At high percentiles the Normal distribution overestimates the actual posterior, while at small percentiles there is an underestimation, except at 0.1-0.01, as is clear also from Figure\,\ref{Fig:priors}. These differences reflect into the posteriors (third column). In particular, the posterior distribution from the EM fit with the Normal GW-informed prior in the majority of cases predicts larger angles with respect to the EM+GW fit. Instead it is the contrary for the EM fit using as prior the actual GW posterior, but the percentage differences are smaller, especially in the tails of the distribution. 

\section{Results for GW170817}

\label{sec:results}

\renewcommand{\arraystretch}{1.5}
\begin{table*}
    \caption{Fit results for GW170817. We report the medians and the 16th-84th percentiles. The angles are quoted in degrees.  $\theta_v$ and $\theta_{\rm JN}$ are related by Eq.\,(\ref{eq:theta_v_JN}) and treated as a single parameter.}
    \begin{tabular}{ccccc} 
% \multicolumn{3}{c}{}} & \multicolumn{3}{c}{}\\
    \hline
    \noalign{\smallskip}
    Parameter & GW-only & EM-only (sine prior) & EM-only (GW-informed prior) & EM+GW \\
    \noalign{\smallskip}
    \hline
    \noalign{\smallskip}
        $\log_{10} E_0$ & & $50.67^{+0.06}_{-0.05}$ & $50.75^{+0.07}_{-0.06}$ & $50.73^{+0.07}_{-0.06}$ \\
        $\theta_c$ [$\degree$] & & $7.0^{+0.4}_{-0.5}$ & $6.1^{+0.3}_{-0.4}$ & $6.2^{+0.4}_{-0.5}$ \\
        $\theta_w$ [$\degree$] & & $54^{+21}_{-20}$ & $49^{+22}_{-19}$ & $51^{+22}_{-20}$ \\
        $\log_{10}n_0$ & & $-3.2^{+0.2}_{-0.2}$ & $-3.6^{+0.1}_{-0.2}$ & $-3.5^{+0.2}_{-0.2}$ \\
        $p$  & & $2.14^{+0.01}_{-0.01}$ & $2.14^{+0.01}_{-0.01}$ & $2.14^{+0.01}_{-0.01}$ \\
        $\log_{10} \epsilon_{\rm e}$ & & $-0.02^{+0.02}_{-0.03}$ & $-0.09^{+0.05}_{-0.05}$ & $-0.07^{+0.04}_{-0.06}$  \\
        $\log_{10} \epsilon_{\rm B}$ & & $-1.8^{+0.1}_{-0.1}$ & $-1.6^{+0.2}_{-0.2}$ & $-1.7^{+0.2}_{-0.2}$ \\
        \hline
        $\theta_{\rm v}$ [$\degree$] & & $38^{+2}_{-2}$ &  $33^{+2}_{-2}$ & $34^{+2}_{-2}$ \\
        $\theta_{\rm JN}$ [$\degree$]&  $151^{+3}_{-3}$ & & & $146^{+2}_{-2}$ \\
        \hline
        $ \mathcal{M}$ & $1.1975^{+0.0001}_{-0.0001}$ & & & $1.1975^{+0.0001}_{-0.0001}$ \\
        $q$ & $0.88^{+0.08}_{-0.10}$ & & & $0.88^{+0.08}_{-0.09}$ \\
        $a_1$ & $0.02^{+0.02}_{-0.01}$ & & & $0.02^{+0.02}_{-0.01}$ \\
        $a_2$ & $0.02^{+0.02}_{-0.01}$ & & & $0.02^{+0.02}_{-0.02}$ \\
        $\theta_1$ [$\degree$]& $82^{+34}_{-35}$ & & & $80^{+33}_{-33}$ \\
        $\theta_2$ [$\degree$]& $83^{+36}_{-36}$ & & & $83^{+36}_{-34}$ \\
        $\phi_{1,2}$ [$\degree$] & $182^{+122}_{-129}$ & & & $177^{+115}_{-115}$ \\
        $\phi_{\rm JL}$ [$\degree$] & $179^{+123}_{-123}$ & & & $177^{+120}_{-121}$ \\
        $\psi$ [$\degree$] & $89^{+61}_{-62}$ & & & $84^{60}_{-54}$ \\
        $\Lambda_1$ & $269^{+346}_{-190}$ & & & $278^{+376}_{-192}$ \\
        $\Lambda_2$ & $425^{+525}_{-295}$ & & & $421^{+528}_{-287}$ \\
    
    \noalign{\smallskip}
    \hline
\end{tabular}
\label{table:results}
\end{table*}
\renewcommand{\arraystretch}{1}

In this section we show the results of the GW-only fit, the afterglow-only fit and the joint fit, obtaining an improvement in the determination of the viewing angle with respect an EM-only fit. The EM fits are performed with \dynesty, and the GW and EM+GW are carried out using \bilby,  and \dynesty, specifically with the dynamic nested sampling method with 2000 livepoints and multiple bounding ellipsoids as bounding strategy. The corner plots in this work are created making use of \textsc{corner} \citep{corner}.

\subsection{GRB 170817A afterglow and GW170817 analyses}

We perform a GW-only fit of GW170817, the results are reported in the first column of Table \ref{table:results}.  As is commonly the case, we do not quote the individual mass components but the chirp mass
\begin{equation}
    \mathcal{M} = \frac{(m_1 m_2)^{3/5}}{(m_1 + m_2)^{1/5}}\,,
\end{equation}
which is the best measured parameter for systems displaying a long inspiral \citep{Finn1993, Blanchet1995, Cutler1994, Poisson1995}, and the mass ratio $q = m_2/m_1 \leq 1$.  The 6 dimensionless spin degrees of freedom are reported as follows: $a_1$ and $a_2$ are the dimensionless spin magnitudes, $\theta_1$ and $\theta_2$ are the tilt angles between the spins and the orbital angular momentum, $\phi_{1,2}$ is the azimuthal angle separating the spin vectors, and $\phi_{JL}$ is the opening angle of the cone of precession about the system angular momentum.  We do not report the time of coalescence in this and other tables, as this is of little interest in the context of our study.
Our results are in agreement with previous works \citet{Abbott2019_properties, Romero-shaw2020}. Our error on the inclination angle $\theta_{\rm JN}$ is about 5 times smaller with respect to \citet{Romero-shaw2020}, who report $\theta_{\rm JN} = 145\degree^{+17\degree}_{-18\degree}$ (68\% credible interval). Our 90\% credible interval for $\theta_{\rm JN}$ is $151\degree^{+6\degree}_{-5\degree}$, which is about 4 times smaller with respect to \citet{Abbott2019_properties}, that report $146\degree^{+25\degree}_{-27\degree}$. This simply stems from the fact that in our analysis we collapse the prior on the luminosity distance, which is correlated to $\theta_{\rm JN}$, by fixing it to a single value. In order to verify this, we repeat the GW fit leaving the luminosity distance free with a uniform-in-volume prior (see the works cited above). The results for the inclination angle are: $\theta_{\rm JN} = 145\degree^{+16\degree}_{-18\degree}$ (68\% credible interval) and $\theta_{\rm JN} = 145\degree^{+25\degree}_{-27\degree}$ (90\% credible interval). These results are very much in agreement with \citet{Abbott2019_properties, Romero-shaw2020}.

\begin{figure*}
    \includegraphics[width=0.7\textwidth]{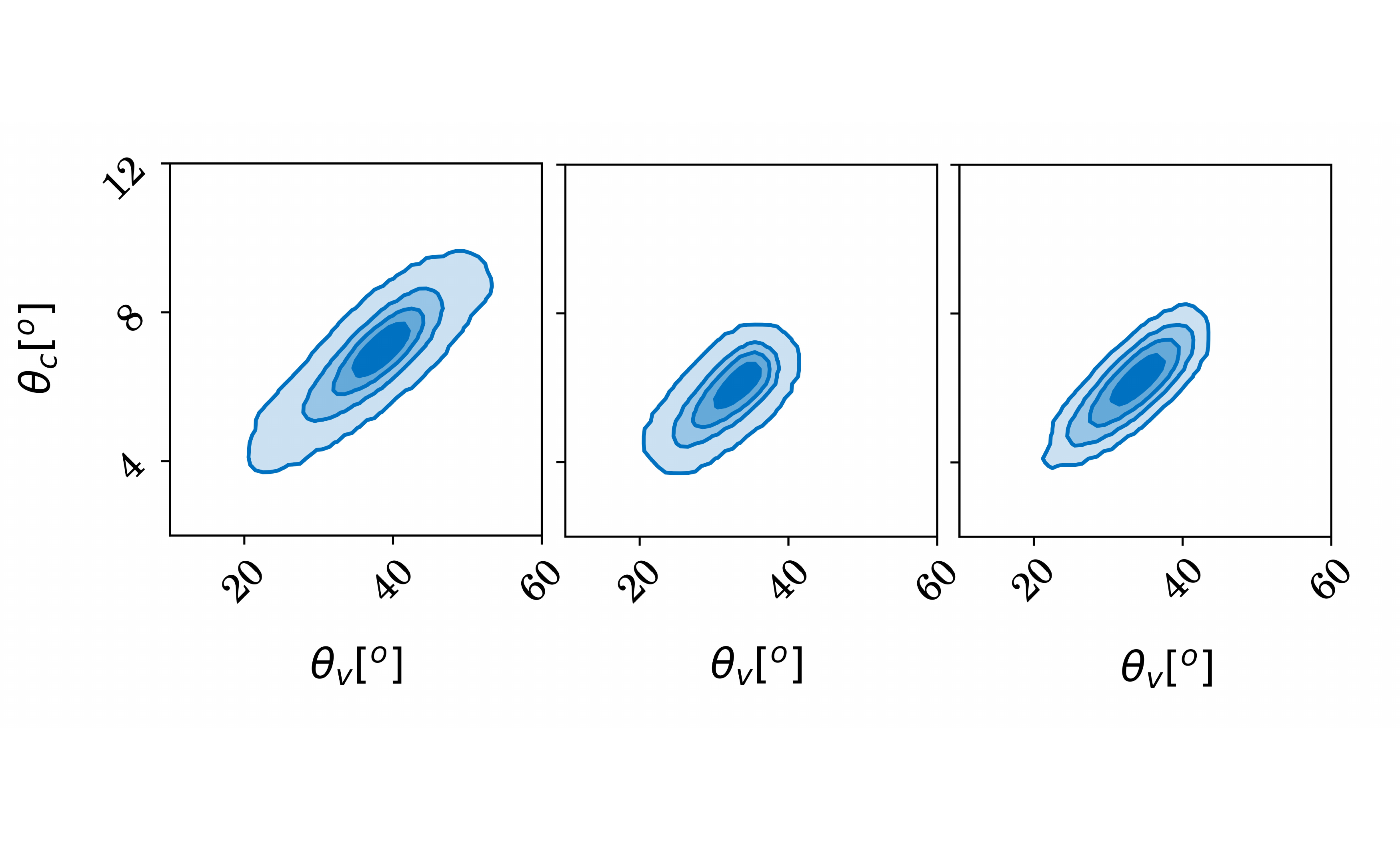}
    \caption{2D distribution of the viewing angle and jet opening angle. From left to right, the result from the  afterglow-only fit with a sine prior on $\theta_v$, afterglow-only fit with GW-informed prior, and the EM+GW fit. The contours represent the 68.2\%, 95.4\%, 99.8\% and 99.98\% probabilities.}
     \label{Fig:comparison}
\end{figure*}

The broad-band afterglow of GW170817 is fitted assuming a Gaussian structure for the jet and using two different priors on the viewing angle: an isotropic prior uniform in the cosine of $\theta_v$ from 0$\degree$ to 90$\degree$ (referred to as ``sine prior'') and the $\theta_{\rm JN}$ posterior distribution of the GW-only analysis (referred to as ``GW-informed prior''), where $\theta_{\rm JN}$ is transformed in $\theta_v$ via Eq.\,(\ref{eq:theta_v_JN}).  The results of the analysis of the afterglow data with two distinct priors are summarized in the second and third column of Table \ref{table:results}. 
The values of the parameters concerning the geometry, the energetics, and the microphysics are in agreement within 1$\sigma$ across the two fits. The viewing angle goes from a median value of 38$\degree$ for the sine prior to 33$\degree$ for the GW-informed prior. These results are in agreement with \citet{troja2021}, who used the same dataset of this work, except for the most recent Chandra data point. Despite the decrease in the viewing angle with the GW-informed prior, $\theta_{\rm v}$ is still in tension with the analysis of the centroid motion in the radio observations by \citet{Mooley2018}, who predict a $\theta_{\rm v} = 20\degree$. This discrepancy is due to the flattening of the late time observations in the X-rays, which is driving the viewing angle toward higher values. As pointed out in \cite{troja2021}, the addition of an extra-component in the flux producing the late time observations could resolve this discrepancy, because it would allow the jet to be narrower and closer to the line of sight. For an extensive discussion see \citet{troja2021, Balasubramanian2021, Hajela2022}. 

\subsection{Joint analysis}

The parameter medians and 16th-84th percentiles of the joint EM+GW fit are collected in the last column of Table\,\ref{table:results}.
In Fig.\,\ref{Fig:comparison} we report the marginalised, 2D posterior probability distribution for the jet opening angle $\theta_c$ and the viewing angle $\theta_v$ in the three cases these parameters are measured within our setups: namely, from left to right, the EM-only fit with sine prior, the EM-only fit with GW-informed prior, and the joint EM+GW fit. While there is a degeneracy in the EM-only fit, including the GW information (either as a prior or as an additional independent dataset) still leaves the degeneracy, but reduce the dispersion. This happens because the EM dataset alone already constrains very well the two angles, especially thanks to the late time detections in the X-rays. The EM fit with the GW informed prior and the EM+GW joint fit are equivalent, as proven in Appendix\,\ref{Sec:Appendix}. The advantage of performing an EM+GW joint fit is the back-reaction of the GRB analysis on the GW parameters, in this case though the GW parameters remain almost the same. This is because the inclination of the binary is degenerate with the luminosity distance, which is, however, fixed in this work. We will explore the case of a free luminosity distance analysis in a future work.

We obtain an inclination of $\theta_{\rm JN} = 146\degree^{+2\degree}_{-2\degree}$ (68\% credible interval), which translates into a viewing angle $\theta_{v} = 34\degree^{+2\degree}_{-2\degree}$. These values are consistent with the ones retrieved with the EM-only fit when using the GW-informed prior.
The error on the common parameter, $\theta_{\rm JN}$, from the joint EM+GW fit is 1.5 times smaller than for the GW-only fit. 

\section{GW170817-like event in O4}
\label{sec:results_O4}

As GW170817 was a nearby and rare event, in this section we repeat our analyses in the scenario in which a GW170817-like source located at 136.5\,Mpc and at 70\,Mpc (redshift $z=0.03$ and $z=0.016$) are observed in the LIGO-Virgo-KAGRA Fourth Observing Run (O4). The increase in distance has the effect of reducing by roughly an order of magnitude and a factor of three the flux of the afterglow respectively. These distances are within the expected horizon of LIGO and Virgo for O4\footnote{\href{https://dcc.ligo.org/LIGO-G2002127/public}{https://dcc.ligo.org/LIGO-G2002127/public}}. Within the latter, the number of binary neutron stars mergers expected in O4 is $\lesssim 10$ \citep{Abbott_ratesO4, Abbott_O3catalogue}. 
We simulated $10^5$ events, producing light curves with the same energetics and microphysics parameters as GW170817 (see Table\,\ref{table:results}, last column), distributed uniformly in volume up to 190 Mpc and with jet axes isotropically distributed. We find that 25\% of events reach peak fluxes above the \textit{Chandra} sensitivity, concluding that the X-ray afterglow of  $\lesssim  3$ GW170817-like binary neutron stars mergers could be detected  during O4. This rate is in agreement with \citet{Wang2021, Petrov2022, Colombo2022, Patricelli2022, Wang2022}.

\subsection{Building the dataset}
\label{sec:datasetO4}

\begin{figure}
    \includegraphics[width=0.5\textwidth]{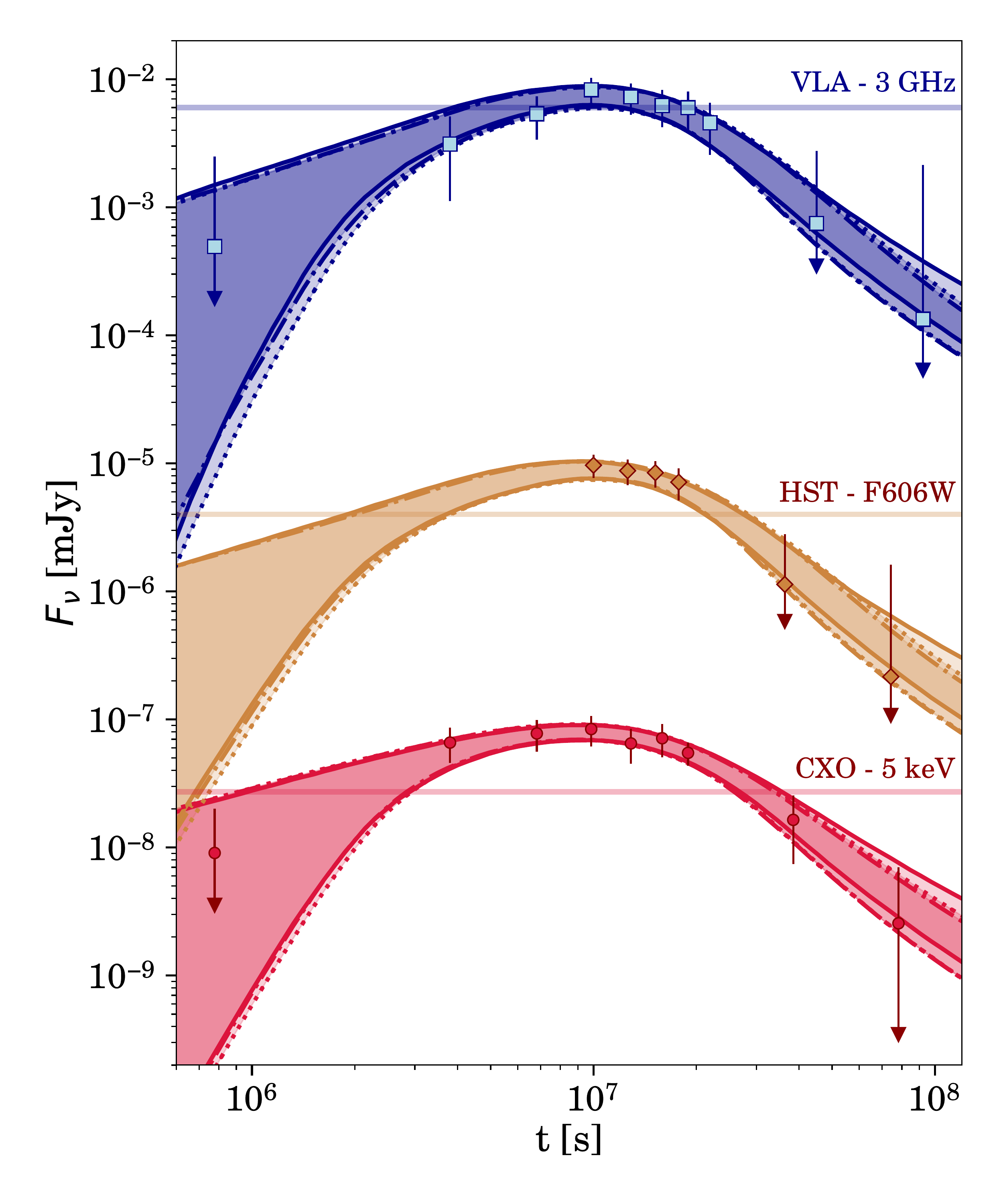}
    \caption{Broad-band afterglow of a GW170817-like event at 136.5\,Mpc: rescaled data and fits.  From bottom to top, red dots are X-ray observations by Chandra at 5\,keV, orange diamonds are the observations by HST, F606W filter, in the optical band, and blue squares are VLA observations 3\,GHz. The arrows indicate that the 1$\sigma$ error reaches 0 flux. The shaded regions and solid, dot-dashed, and dotted lines represent the 68\% uncertainty regions of the models envelope from the EM-only (sine prior), EM-only (GW prior), and EM+GW fits of the afterglow, respectively. Note that the EM with GW prior and EM+GW fits have almost identical uncertainty regions. The horizontal shaded lines represent the sensitivity of \textit{Chandra} (red, $2.7\times10^{-8}$mJy, $3\sigma$), \textit{HST} (orange,  $4\times10^{-6}$mJy, $3\sigma$) and VLA (blue,  $6\times10^{-3}$mJy, $3\sigma$).}
     \label{Fig:lightcurve_O4}
\end{figure}

In the EM sector, we rescale the afterglow of GW170817 for the luminosity distance of 136.5\,Mpc (or 70\,Mpc). In the case of 136.5\,Mpc, for example, this reduces the flux of about one order of magnitude, as seen by comparing Figs.\,\ref{Fig:lightcurve_O4} and \ref{Fig:lightcurve_gw170817}. We adopt a minimal and conservative observing strategy: starting from 9 days, which is the first detection in the X-rays of GW170817, we place an observation every 4 weeks until the peak. After that, we put equally spaced (in log space) observations in the following time decade. The details for each band are as follows.

\begin{itemize}
\item Radio --- We consider only the (rescaled) VLA observations at 3\,GHz. We assign an error of $2\mu$Jy, which is the root-mean-square error estimated for more than 3\,hr of observations \citep{Smol2017}. In this respect, a radio observation of GW170817 with 6\% uncertainty has 30\% uncertainty at 136.5 Mpc.

\item Optical --- For the \textit{HST} optical observations, and, in particular, their uncertainties, we study the contributions of source, background, readnoise and noise from the dark current \citep{DesjardinsLucas2019}. The \textit{HST} readnoise and the dark current are negligible with respect to the source and sky contributions. The source weights 10--20\% of the error, so the error is dominated by the sky (background). We scale with distance only the source part of the error, then we estimate the error contribution of the sky assuming an exposure time of 5\,ks, and we sum in quadrature the two components to have the final error on the optical fluxes. In this case, the about 20\% errors on GW170817 become 30\% errors for O4 observations at 136.5\,Mpc.
\item X-ray --- We use the software \xspec\footnote{\href{https://heasarc.gsfc.nasa.gov/xanadu/xspec/}{Xspec Home Page}} \citep{xspec} to generate the X-ray observations taking into account the \textit{Chandra} response and background\footnote{\href{https://cxc.cfa.harvard.edu/caldb/prop_plan/imaging/}{\textit{Chandra} X-ray Observatory}}. The GW170817 X-ray observations with almost 10\% uncertainty has about 30\% uncertainty in the re-scaled event at 136.5\,Mpc.
\end{itemize}

Finally, in order to make this result valid for real observations, we check the sensitivity of the instruments. 
The Chandra sensitivity for 100\,ks exposure is about $2.7\times10^{-8} \rm mJy$ (3$\sigma$ at 5\,keV). 
For VLA, we use a 3$\sigma$ sensitivity of $6\mu$Jy (see above). 
For HST, we assume a $3\sigma$ limiting magnitude of about 30 ABmag, which could be reached stacking different images together. For this reason we chose an exposure time of 5\,ks when treating the errors. Figure \ref{Fig:lightcurve_O4} depicts the final EM dataset for the putative, GW170817-like O4 event at 136.5\,Mpc. The sensitivity of each instrument is represented by a horizontal line. In this case, all three instruments can detect only the peak of the afterglow. We keep also the data points below the sensitivities, as they will play a major role in constraining the jet opening angle and the viewing angle. 
There is no data point at 9 days in the optical band because at that time the kilonova dominates the emission. 

In the GW sector, we use \bilby to produce a GW signal from a source with the same features as GW170817, but located at 136.5\,Mpc (70\,Mpc), and inject it in the nominal O4 noise curves\footnote{\href{https://dcc.ligo.org/LIGO-T2000012/public}{https://dcc.ligo.org/LIGO-T2000012/public}} for the LIGO-Livingston, LIGO-Hanford and Virgo detectors. We do not include the KAGRA detector in this analysis, since a binary neutron star at $\sim 140 (70) \,$Mpc would be beyond its expected sensitivity range. This simulated event has a network signal-to-noise ratio of 26.3 (50), as opposed to the GW170817 network signal-to-noise ratio of 32.4.

In order to validate this GW+EM injection setup, we simulate an event similar to GW170817 with the O2 interferometers noise curves. The injected GW signal has the same parameters of GW170817, including the luminosity distance (see Table\,\ref{table:results}, first column). The EM dataset is composed of the same detection times and flux errors of the GW170817 broad-band afterglow, but we adopt the flux predicted by the model (and not the real, detected, one).
Jointly fitting this simulated EM+GW dataset leads to the same results of the real GW170817, so our injection set up can be safely used for the following analysis of the other events.

\subsection{Results for a distance of 136.5\,Mpc}
\label{subsec:136Mpc}

\begin{figure}
    \includegraphics[width=0.5\textwidth]{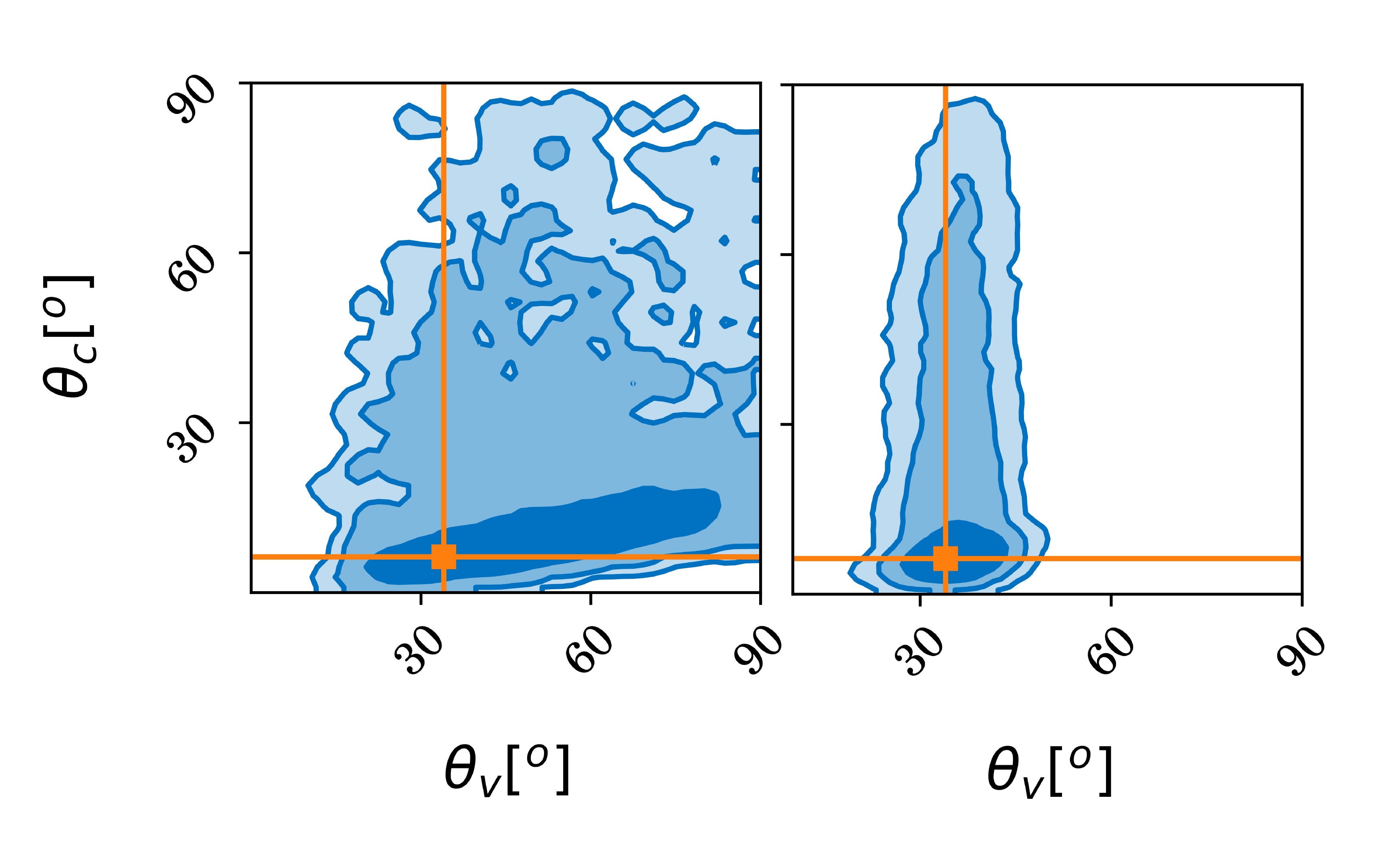}
    \caption{2D distribution of the viewing angle and jet opening angle for a GW170817-like event located at 136.5 Mpc. On the left, the result from the afterglow-only fit with a sine prior on $\theta_v$, on the right the EM+GW fit. The contours represent the  68.2\%, 95.4\% and 99.8\% probabilities. The orange squares represents the values used to simulate the observations (see Table \ref{table:results}, last column) for $\theta_c$ and $\theta_v$.}
     \label{Fig:comparison_o4}
\end{figure}

We fit the EM and GW datasets described above using \bilby and \dynesty (dynamic nested sampling method with 2000 livepoints and multiple bounding ellipsoids). For this event, in the contour plots we will show only the results from the EM-only and the EM+GW joint fit, being the EM fit with GW-informed prior equal to the latter, as already demonstrated for GW170817 (Section \ref{sec:results}). 
The results are reported in Table \ref{table:results_O4}, Fig.\,\ref{Fig:lightcurve_O4}, and Fig.\,\ref{Fig:comparison_o4}.  The table lists 68\% credible intervals for the parameters involved in the GW-only, EM-only, and joint fit.  Fig.\,\ref{Fig:lightcurve_O4} shows the 68\% uncertainty regions of the fits to the afterglow data; this involves three analyses (afterglow-only fit with sine-prior for $\theta_v$, afterglow-only fit with GW-informed-prior for $\theta_v$, and the joint EM+GW fit) performed over three EM bands (radio, optical, and X-ray).  Finally, Fig.\,\ref{Fig:comparison_o4} displays the contour plots of the marginalized 2D posterior density distributions for $\theta_v$ and $\theta_c$ obtained for the analyses involving the EM dataset and the EM+GW fit.

\renewcommand{\arraystretch}{1.5}
\begin{table*}
    \caption{Fit results for a GW170817-like event at 136.5\,Mpc. We report the medians and the 16th-84th percentiles. The angles are quoted in degrees.  $\theta_v$ and $\theta_{\rm JN}$ are related by Eq.\,(\ref{eq:theta_v_JN}) and treated as a single parameter. In the first column there are the values injected in the GW signal.}
    \begin{tabular}{cccccc} 
% \multicolumn{3}{c}{}} & \multicolumn{3}{c}{}\\
    \hline
    \noalign{\smallskip}
    Parameter & Injection & GW-only & EM-only (sine prior) & EM-only (GW-informed prior) & EM+GW \\
    \noalign{\smallskip}
    \hline
    \noalign{\smallskip}
        $\log_{10} E_0$ &  & & $51.6^{+1.0}_{-0.7}$ & $52.0^{+1.2}_{-1.0}$ & $52.1^{+1.1}_{-1.0}$ \\
        $\theta_c$ [$\degree$] &  & & $11^{+13}_{-4}$ & $7^{+13}_{-1}$ & $7^{+15}_{-2}$ \\
        $\theta_w$ [$\degree$] &  & & $39^{+28}_{-20}$ & $36^{+31}_{-20}$ & $29^{+34}_{-14}$ \\
        $\log_{10}n_0$ &  & & $-1.0^{+1.0}_{-1.4}$ & $-2.0^{+1.2}_{-1.0}$ & $-2.0^{+1.1}_{-1.1}$ \\
        $p$  &  & & $2.12^{+0.03}_{-0.04}$ & $2.12^{+0.04}_{-0.04}$ & $2.12^{+0.04}_{-0.04}$ \\
        $\log_{10} \epsilon_{\rm e}$ &  & & $-0.9^{+0.6}_{-0.9}$ & $-1.2^{+0.9}_{-1.2}$ & $-1.2^{+0.9}_{-1.2}$  \\
        $\log_{10} \epsilon_{\rm B}$ &  & & $-3.7^{+1.0}_{-0.8}$ & $-3.3^{+1.2}_{-1.1}$ & $-3.4^{+1.3}_{-1.1}$ \\
        \hline
        $\theta_{\rm v}$ [$\degree$] & & & $50^{+19}_{-16}$ &  $36^{+3}_{-3}$ & $36^{+3}_{-3}$ \\
        $\theta_{\rm JN}$ [$\degree$] & 151 &  $150^{+6}_{-6}$ & &  & $144^{+3}_{-3}$ \\
        \hline
        $ \mathcal{M}$ & 1.1975 & $1.1975^{+0.0001}_{-0.0001}$ & & & $1.1975^{+0.0001}_{-0.0001}$ \\
        $q$ & 0.88 & $0.89^{+0.07}_{-0.09}$ & & & $0.89^{+0.07}_{-0.09}$ \\
        $a_1$ & 0.02 & $0.02^{+0.02}_{-0.01}$ & & & $0.02^{+0.02}_{-0.01}$ \\
        $a_2$ & 0.02 & $0.02^{+0.02}_{-0.01}$ & & & $0.02^{+0.02}_{-0.02}$ \\
        $\theta_1$ [$\degree$] & 82 & $80^{+37}_{-33}$ & & & $78^{+34}_{-34}$ \\
        $\theta_2$ [$\degree$] & 83 & $83^{+35}_{-35}$ & & & $83^{+34}_{-34}$ \\
        $\phi_{1,2}$ [$\degree$] & 182 & $179^{+123}_{-123}$ & & & $169^{+124}_{-111}$ \\
        $\phi_{\rm JL}$ [$\degree$] & 179 & $179^{+118}_{-119}$ & & & $182^{+114}_{-123}$ \\
        $\psi$ [$\degree$] & 89 & $89^{+56}_{-65}$ & & & $88^{+56}_{-63}$ \\
        $\Lambda_1$ & 260 & $148^{+182}_{-107}$ & & & $148^{+191}_{-106}$ \\
        $\Lambda_2$ & 430 & $221^{+279}_{-155}$ & & & $223^{+261}_{-157}$ \\
    
    \noalign{\smallskip}
    \hline
\end{tabular}
\label{table:results_O4}
\end{table*}
\renewcommand{\arraystretch}{1}

The GW-only fit retrieves posteriors (second column in Table \ref{table:results_O4}) consistent with the injected values (first column in Table \ref{table:results_O4}). 
The signal-to-noise ratio of our putative event is not very different from
that of GW170817, it is less than the GW170817 signal-to-noise ratio by a factor of $\sim1.2$, so also the posteriors are very similar to the case of GW170817 discussed in the previous sections, with the exception of the tidal parameters, where the errors are reduced by a factor of about 2.
This is due to the fact that in this case we are not dealing with real data, and are therefore not subject to a specific noise realization, believed to be at the origin of the bimodal distribution for the tidal deformability in the case of GW170817 (see the discussion and Fig.\,11 of \citealt{Abbott2019_properties}).  In this case, following Eq.\,(\ref{eq:theta_v_JN}), the posterior distribution for $\theta_{\rm JN}$ translates into a viewing angle of $30\degree^{+6\degree}_{-6\degree}$ (68\% credible interval around the median).

Overall, we find that the uncertainties on the EM parameters are larger with respect to the analysis of the afterglow of GW170817 (see Table \ref{table:results}). This is due to the fact that the (simulated) decrease in EM flux drives several detections below the sensitivity of the instruments.
However, the <3$\sigma$ data points help to constrain the EM parameters, especially the jet opening angle. From Table\,\ref{table:results_O4} it is clear that in the EM case both angles are unconstrained, while including the GW domain helps constraining the viewing angle, but not the jet opening angle. To show the relevance of the <3$\sigma$ data points we excluded them from the analysis, leading to medians and 16th-84th percentiles:  $\theta_v=64\degree^{+21\degree}_{-17\degree}$ and $\theta_c=45\degree^{+27\degree}_{-25\degree}$ in the case of an EM-only fit;  $\theta_v=36\degree^{+3\degree}_{-3\degree}$ and $\theta_c=35\degree^{+31\degree}_{-26\degree}$ in the EM+GW case. The absence of these data points mainly affects the jet opening angle, which in the EM+GW (EM) case has an 84th percentile of 66\degree (72\degree), these values are 3 times larger than the ones written in Table\,\ref{table:results_O4} and estimated including the <3$\sigma$ data points in the analysis.

In Table \ref{table:results_O4}, for the EM fit with a sine prior, we find a viewing angle of $50\degree^{+19\degree}_{-16\degree}$, which is compatible within 1$\sigma$ with the injected value from GW170817. The credible interval, however, is still larger than the one for GW170817, indicating that an afterglow-only analysis of an event of this kind would not provide a stringent measurement of the viewing angle. Also the jet opening angle of $7\degree^{+15\degree}_{-2\degree}$ is in agreement with the GW170817 result, but the uncertainty is large. The situation improves when we use the $\theta_{\rm JN}$ posterior distribution from the GW-only analysis, transformed into $\theta_v$, as the GW-informed prior for the EM fit. In this case, we find a viewing angle of $36\degree^{+3\degree}_{-3\degree}$, which is in agreement with the GW170817 results presented in Sec.\,\ref{sec:results}. This improvement follows directly from the use of a more informative prior $\theta_v$. Despite this improvement in the measure of $\theta_v$, the jet opening angle uncertainty is still large because the rising part of the afterglow is characterized only by <3$\sigma$ data points. Also in this case, all the parameters regarding the microphysics and the energetics of the jet are in agreement with the respective GW170817 values within 2$\sigma$. 

The case of the EM+GW joint analysis is very similar to the afterglow-only analysis with the GW-informed prior, as expected. The uncertainty on the viewing angle at 1$\sigma$, $36\degree\pm3\degree$ ($^{+6\degree}_{-5\degree}$, 90\% credible interval) is about 6 times smaller when compared to the EM-only analysis with sine prior, $50\degree^{+19\degree}_{-16\degree}$($^{+30\degree}_{-24\degree}$, 90\% credible interval), and 2 times smaller than the GW-only analysis $150\degree\pm6\degree$( $\pm7\degree$, 90\% credible interval). 

In this scenario $\theta_c$ at the 2$\sigma$ level remains unconstrained by the EM data (see Fig.\,\ref{Fig:comparison_o4}), for the reasons discussed previously. The large errors in $\theta_c$ of all three analyses translate into the large shaded regions in Fig.\,\ref{Fig:lightcurve_O4}, which represent the 68\% uncertainty regions. The latter concentrate in the rise of the afterglow, furthermore, the EM analysis with sine prior (light shaded region) yields to higher flux values at late times, due to the large viewing angle.

For this event, at 2$\sigma$ there are cases where $\theta_v < \theta_c$ (see for example Fig.\,\ref{Fig:comparison_o4}), however, those same cases have also $\theta_w < \theta_c$. This means that the jet ends at $\theta_w$ and the angle relevant to discriminate between the on-axis or off-axis case is $\theta_w$. When this is true, the jet is a top hat, which does not have wings, but just a uniform core ending at $\theta_w$. In these cases, the jet is still off-axis if $\theta_v > \theta_w$, but the structure changed. The parameter estimation returns a top hat geometry ($\theta_w < \theta_c$) with a 23\% posterior probability for the EM+GW fit, a 45\% posterior probability for the EM fit with GW-informed prior and a 43\% posterior probability for the EM-only fit.

\begin{figure}
    \includegraphics[width=0.5\textwidth]{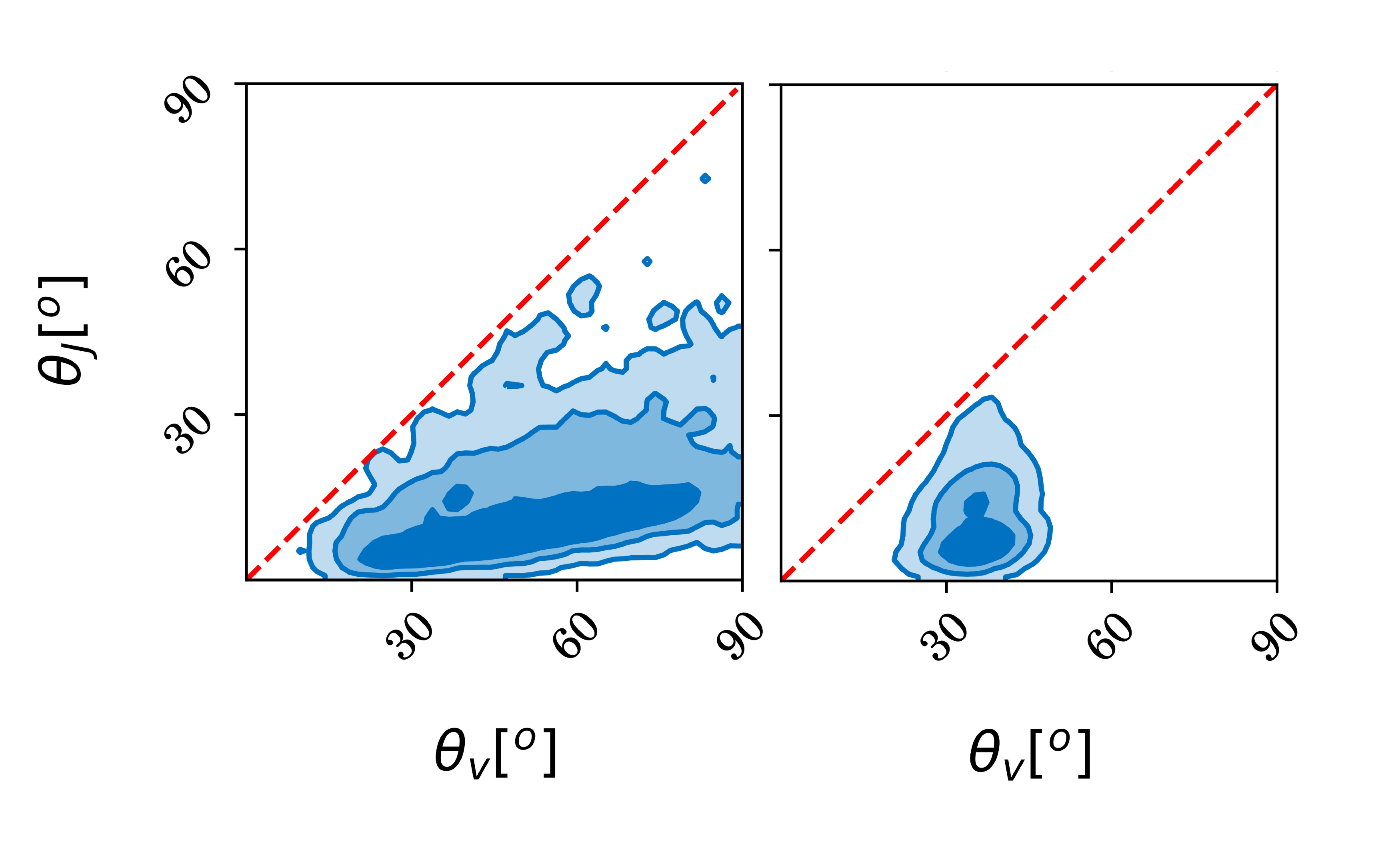}
    \caption{2D distribution of the viewing angle and $\theta_J$, the jet width, for a GW170817-like event located at 136.5 Mpc. On the left, the result from the afterglow-only fit with a sine prior on $\theta_v$, on the right the EM+GW fit. The contours represent the 68.2\%, 95.4\% and 99.8\% probabilities. The red dashed line represents the $\theta_J=\theta_v$ line.}
     \label{Fig:comparison_o4_thetaJ}
\end{figure}

In order to discern between the on-axis and off-axis cases in the samples, we define a new parameter, $\theta_J$, the total width of the jet, that assumes the minimum value between $\theta_w$ and $\theta_c$. The 2D posterior distribution of $\theta_J$ and $\theta_v$ is represented in Fig.\,\ref{Fig:comparison_o4_thetaJ}, where the red dashed line represents the $\theta_J=\theta_v$ line. It is clear that the fit predicts an off-axis observer, with a reduced scatter in the viewing angle when introducing the GW domain. The medians and 16th-84th percentiles for $\theta_J$ are: in the case of the EM-only fit $\theta_J=11\degree^{+7\degree}_{-4\degree}$, in the case of the EM fit with GW-informed prior $\theta_J=7\degree^{+6\degree}_{-1\degree}$ and in the EM+GW case $\theta_J=7\degree^{+6\degree}_{-2\degree}$. 

This analysis shows that, in the case of a large distance event, the data are consistent within 1$\sigma$ with a Gaussian geometry, but we cannot completely rule out a top hat geometry.

\subsection{Results for a distance of 70\,Mpc}
\label{subsec:70Mpc}

\begin{figure}
    \includegraphics[width=0.5\textwidth]{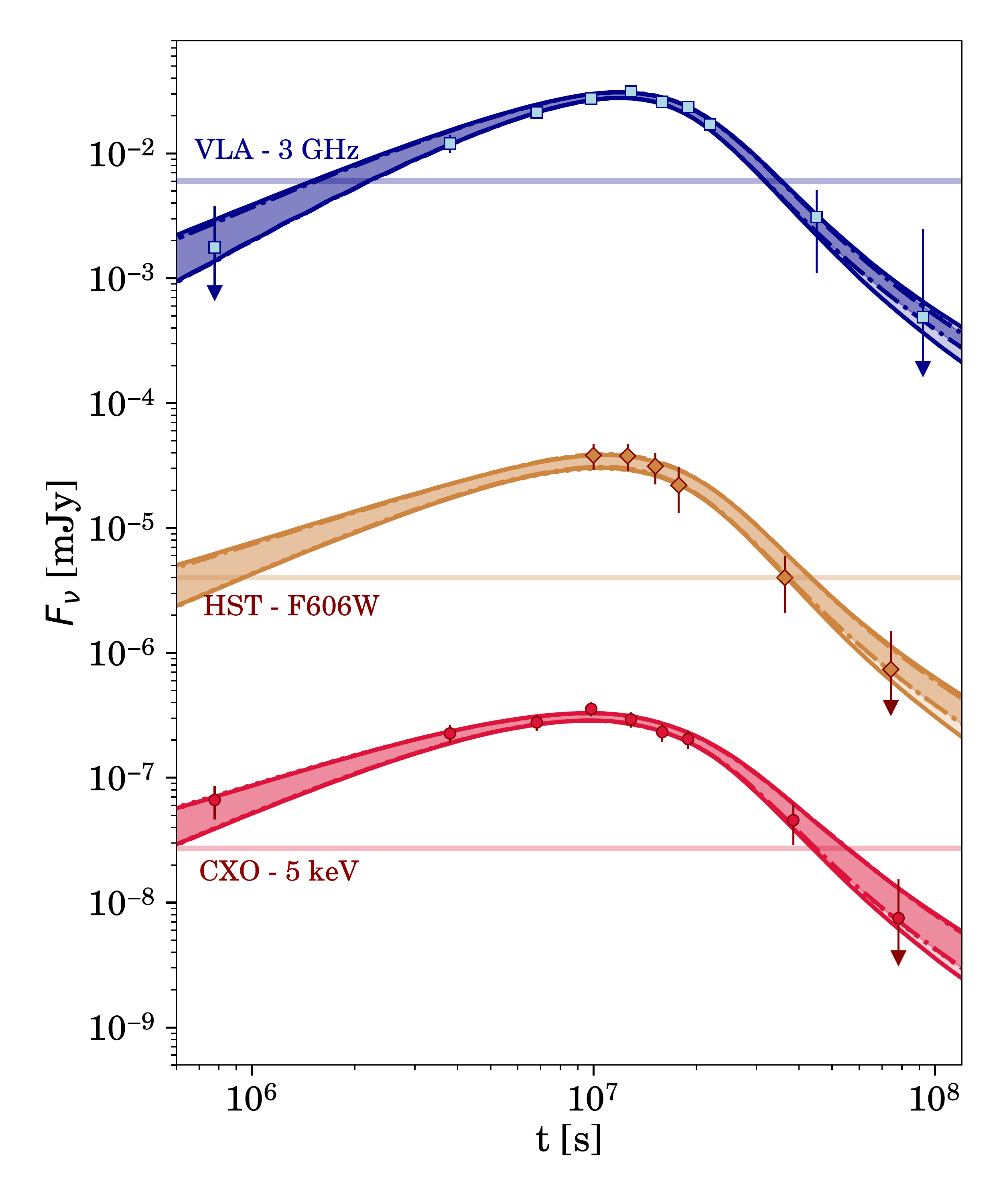}
    \caption{Broad-band afterglow of the GW170817-like event at 70\,Mpc: rescaled data and fits.  From bottom to top, red dots are X-ray observations by Chandra at 5\,keV, orange diamonds are the observations by HST, F606W filter, in the optical band, and blue squares are VLA observations 3\,GHz. The arrows indicate that the 1$\sigma$ error reaches 0 flux. The horizontal shaded lines represent the sensitivity of \textit{Chandra} (red, $2.7\times10^{-8}$mJy, $3\sigma$), \textit{HST} (orange,  $4\times10^{-6}$mJy, $3\sigma$) and VLA (blue,  $6\times10^{-3}$mJy, $3\sigma$). The shaded regions and solid, dot-dashed, and dotted lines represent the 68\% uncertainty regions of the models envelope from the EM-only (sine prior), EM-only (GW prior), and EM+GW fits of the afterglow, respectively. Note that the EM with GW prior and EM+GW fits have identical uncertainty regions.}
     \label{Fig:lightcurve_O4_70}
\end{figure}

In order to study how the degeneracy between the jet opening angle and the viewing angle changes with distance, we also simulate another GW170817-like event, located at the intermediate distance of 70 Mpc ($z\sim0.016$). This increase in distance reduces by roughly a factor of three the flux of the afterglow. The EM and GW datasets are generated as explained in Section \ref{sec:datasetO4}. For this event, in the radio band the first and the last two data points, in the optical and X-ray the last data point are below the sensitivities, the light curve is represented in Fig.\,\ref{Fig:lightcurve_O4_70}. It is clear that the rising phase at 70\,Mpc is better measured than at 136.5\,Mpc.

Notwithstanding the larger distance, the GW event has an improved signal-to-noise ratio by a 1.5 factor with respect to GW170817 thanks to the improved interferometer performances predicted in O4. This explains the smaller errors in $\theta_{\rm JN}$, $\mathcal{M}$ and $q$ (see Table\,\ref{table:results_O4_70}). The considerations done in Section\,\ref{subsec:136Mpc} about the tidal parameters are valid also in this case. The results of the fitting are listed in Table\,\ref{table:results_O4_70}.

\begin{figure*}
    \includegraphics[width=0.6\textwidth]{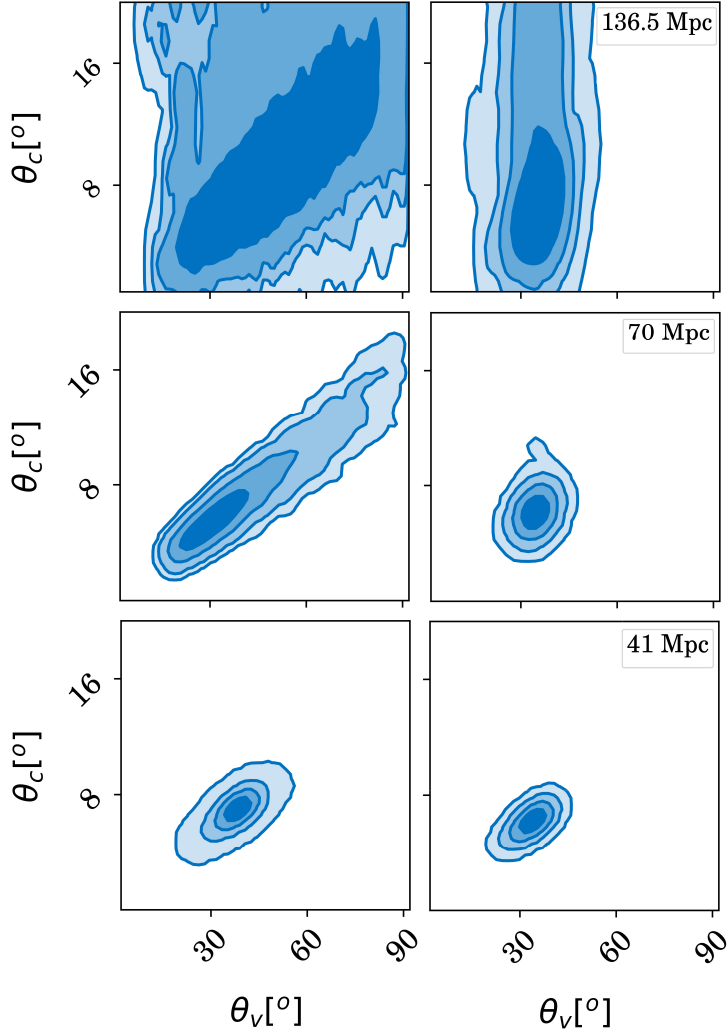}
    \caption{2D posterior distributions of the EM-only fit (first column) and the joint EM+GW fit (second column). In the first row the event at 136.5\,Mpc is represented, in the second row the simulated event at 70\,Mpc and in the third row the GW170817 event. The contour lines represent the 68.2\%, 95.4\%, 99.8\% and 99.98\% probabilities. The posteriors shown in the first row are the same as Fig.\,\ref{Fig:comparison_o4}, with a different scale for $\theta_c$. It is to be noted that the event at 41\,Mpc (GW170817) occurred in the second observing run (O2), while the other events are simulated using the O4 noise curves of the LIGO and Virgo interferometers.}
     \label{Fig:concl}
\end{figure*}

All the EM parameters are in agreement with GW170817 within 2$\sigma$. The reduced uncertainty on the parameters with respect to the 136.5\,Mpc event translates into narrower 68\% uncertainty regions, represented in Fig.\,\ref{Fig:lightcurve_O4_70}. The $\theta_v - \theta_c$ posterior distribution is shown in Fig.\,\ref{Fig:concl}, second row. It is clear that the jet opening angle and the viewing angle are better constrained than the large distance event in all fits, and $\theta_v > \theta_c$ always, as in the case of GW170817. However, there is still a strong degeneracy in the EM-only fit, which is completely broken including the GW domain (leading also to slightly smaller EM+GW shaded regions in Fig.\,\ref{Fig:lightcurve_O4_70} with respect to the EM-only fit). This is not happening for GW170817 because the presence of the correlation at 41\,Mpc in the EM+GW fit is due to the small errors on the angles from the EM modelling, that do not leave much margin of improvement. At this distance, the jet opening angle is very well constrained, with an about 10\% error in the EM+GW case, with respect to the large distance event (more than 100\% error). 

\renewcommand{\arraystretch}{1.5}
\begin{table*}
    \caption{Fit results for a GW170817-like event at 70\,Mpc. We report the medians and the 16th-84th percentiles. The angles are quoted in degrees.  $\theta_v$ and $\theta_{\rm JN}$ are related by Eq.\,(\ref{eq:theta_v_JN}) and treated as a single parameter. In the first column there are the values injected in the GW signal.}
    \begin{tabular}{cccccc} 
% \multicolumn{3}{c}{}} & \multicolumn{3}{c}{}\\
    \hline
    \noalign{\smallskip}
    Parameter & Injection & GW-only & EM-only (sine prior) & EM-only (GW-informed prior) & EM+GW \\
    \noalign{\smallskip}
    \hline
    \noalign{\smallskip}
        $\log_{10} E_0$ &  & & $51.1^{+1.4}_{-0.3}$ & $51.0^{+1.2}_{-0.2}$ & $51.0^{+1.0}_{-0.2}$ \\
        $\theta_c$ [$\degree$] &  & & $6^{+2}_{-1}$ & $6.1^{+0.6}_{-0.6}$ & $6.0^{+0.6}_{-0.6}$ \\
        $\theta_w$ [$\degree$] &  & & $53^{+21}_{-22}$ & $54^{+21}_{-21}$ & $52^{+22}_{-22}$ \\
        $\log_{10}n_0$ &  & & $-3.2^{+1.5}_{-0.8}$ & $-3.3^{+1.2}_{-0.2}$ & $-3.2^{+1.0}_{-0.3}$ \\
        $p$  &  & & $2.14^{+0.03}_{-0.03}$ & $2.14^{+0.02}_{-0.03}$ & $2.14^{+0.02}_{-0.03}$ \\
        $\log_{10} \epsilon_{\rm e}$ &  & & $-0.2^{+0.2}_{-1.3}$ & $-0.2^{+0.1}_{-1.1}$ & $-0.2^{+0.1}_{-0.8}$  \\
        $\log_{10} \epsilon_{\rm B}$ &  & & $-2.1^{+0.6}_{-1.5}$ & $-2.0^{+0.4}_{-1.2}$ & $-2.1^{+0.4}_{-1.1}$ \\
        \hline
        $\theta_{\rm v}$ [$\degree$] & & & $32^{+9}_{-7}$ &  $34^{+2}_{-2}$ & $35^{+2}_{-2}$ \\
        $\theta_{\rm JN}$ [$\degree$] & 151 &  $150^{+2}_{-2}$ & &  & $145^{+2}_{-2}$ \\
        \hline
        $ \mathcal{M}$ & 1.1975 & $1.19747^{+0.00004}_{-0.00003}$ & & & $1.19747^{+0.0004}_{-0.0003}$ \\
        $q$ & 0.88 & $0.89^{+0.07}_{-0.08}$ & & & $0.89^{+0.07}_{-0.08}$ \\
        $a_1$ & 0.02 & $0.02^{+0.02}_{-0.01}$ & & & $0.02^{+0.02}_{-0.01}$ \\
        $a_2$ & 0.02 & $0.02^{+0.02}_{-0.01}$ & & & $0.02^{+0.02}_{-0.02}$ \\
        $\theta_1$ [$\degree$] & 82 & $83^{+32}_{-33}$ & & & $82^{+33}_{-33}$ \\
        $\theta_2$ [$\degree$] & 83 & $84^{+35}_{-35}$ & & & $84^{+34}_{-35}$ \\
        $\phi_{1,2}$ [$\degree$] & 182 & $179^{+124}_{-124}$ & & & $179^{+124}_{-122}$ \\
        $\phi_{\rm JL}$ [$\degree$] & 179 & $180^{+122}_{-126}$ & & & $182^{+120}_{-123}$ \\
        $\psi$ [$\degree$] & 89 & $87^{+54}_{-64}$ & & & $88^{+55}_{-63}$ \\
        $\Lambda_1$ & 260 & $161^{+158}_{-110}$ & & & $162^{+151}_{-109}$ \\
        $\Lambda_2$ & 430 & $236^{+224}_{-163}$ & & & $237^{+213}_{-159}$ \\
    
    \noalign{\smallskip}
    \hline
\end{tabular}
\label{table:results_O4_70}
\end{table*}
\renewcommand{\arraystretch}{1}

\section{Conclusions}
\label{sec:conclusions}

In this work we exploit Bayesian analysis to process the GW and EM afterglow data of the multi-messenger event GW170817, and apply the same methodology to two GW170817-like events originating at greater distances and occurring in O4. We keep the luminosity distance fixed, as the usual procedure in the analysis of afterglow properties and we assume a Gaussian structure for the jet.
We investigate whether a joint fit of EM and GW data brings improvement with respect to fitting those datasets independently and/or with informed priors. We also show that the joint fit and an EM fit with a GW-informed prior are equivalent, as long as the actual posterior of the GW analysis is used as prior for the EM analysis. On the contrary, using an approximation of the GW posterior as prior can lead to differences in the resulting posteriors. 

The GW and EM domains share a common parameter, namely the inclination angle $\theta_{\rm JN}$ (from the GW side) and the viewing angle  $\theta_v$ (from the EM side), which are related by Eq.\,(\ref{eq:theta_v_JN}).
In the EM model used to process the data, the jet opening angle $\theta_c$ and $\theta_v$ are correlated: we show that a joint analysis of the two messengers can ease or break this degeneracy and provide more accurate measurements of $\theta_v$ than those obtained analysing the GW or the EM data alone.

\begin{figure}
    \includegraphics[width=0.5\textwidth]{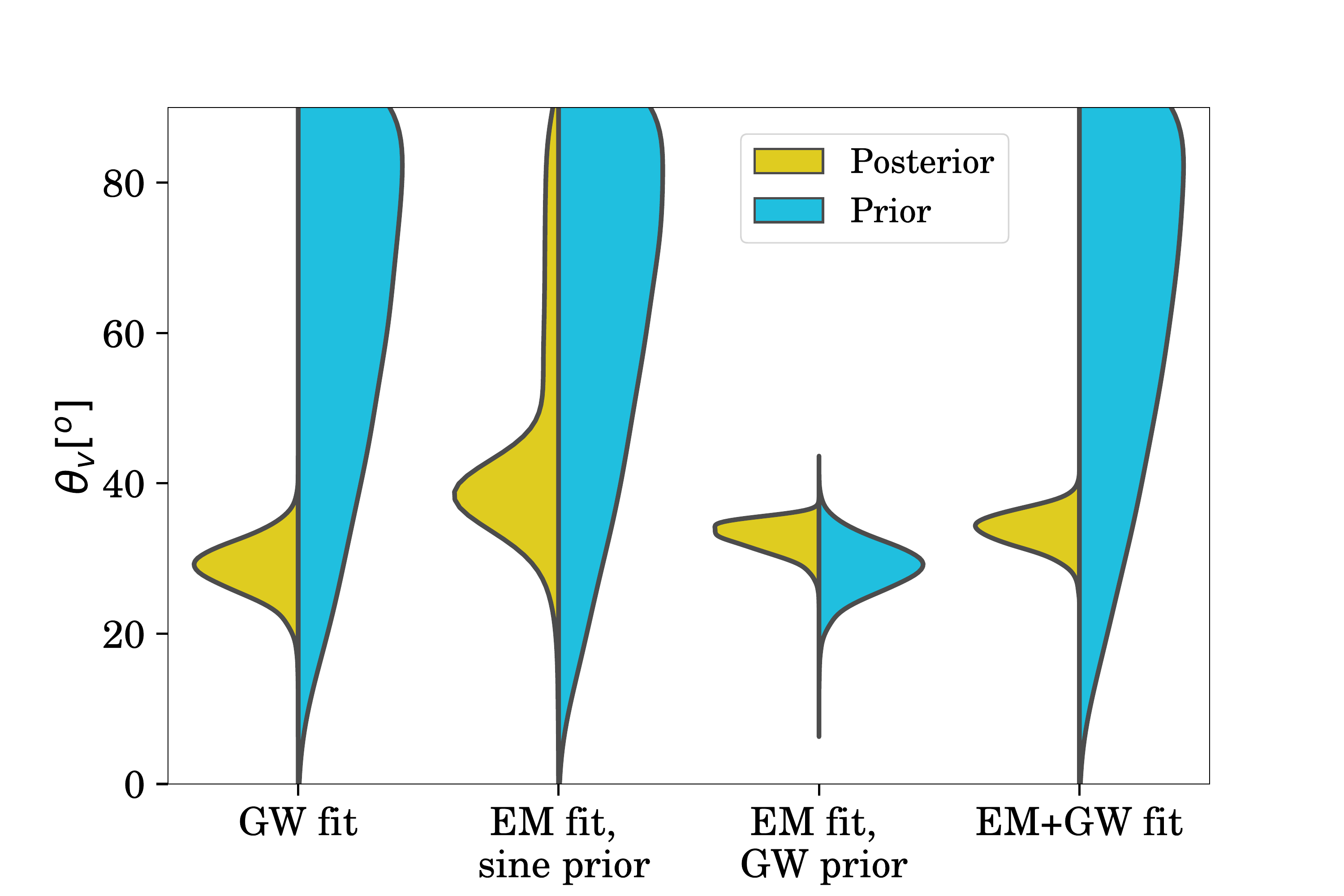}
    \includegraphics[width=0.5\textwidth]{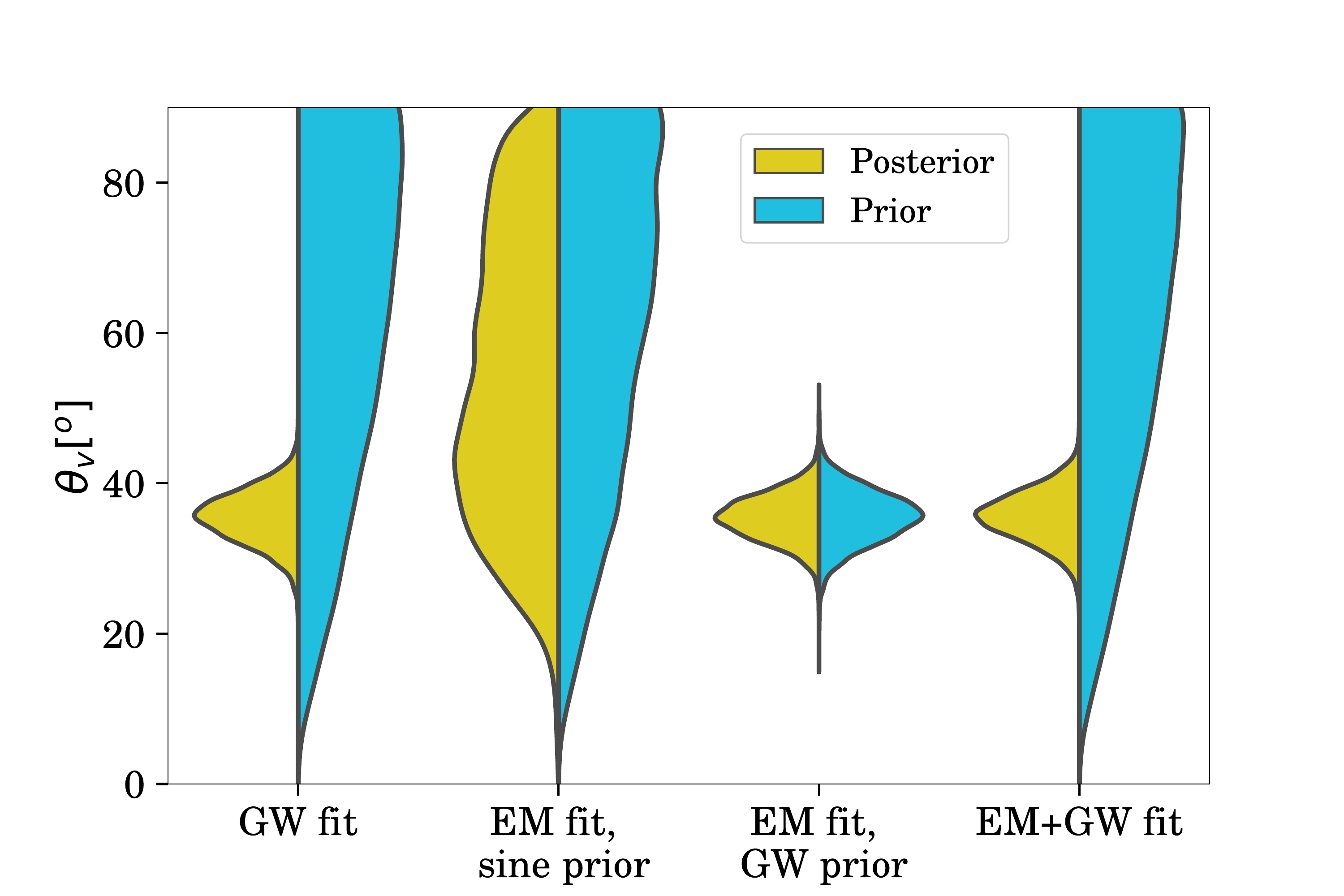}
    
    \caption{Prior (cyan) and posterior (yellow) distributions of the viewing angle $\theta_v$ for the analyses carried out on GW170817 data (top panel) and a simulated GW170817-like event in O4 (bottom panel). In both panels, the cyan distributions represent a distribution uniform in the cosine of the angle with the exception of the third case from the left, where the prior distribution corresponds to the posterior from the GW analysis (represented in yellow in the first case from the left).}
     \label{Fig:prior_posterior_thetav}
\end{figure}

A comprehensive summary of all our results for the measurement of $\theta_v$ in the extreme cases of GW170817 and of the large distance event 
at 136.5\,Mpc is shown in Fig.\,\ref{Fig:prior_posterior_thetav}, where, for both scenarios, we compare the prior on $\theta_v$ to the posterior on the same quantity obtained by: 1) processing only the GW data  (and applying Eq.\,(\ref{eq:theta_v_JN}) to convert the $\theta_{\rm JN}$ measurement), 2) processing only the EM data when assuming a uniform prior on the cosine of $\theta_v$ (``sine prior''), 3) processing separately the GW and the EM data, using the $\theta_{\rm JN}$ posterior distribution as prior on $\theta_v$, and 4) processing both datasets simultaneously.

We see that in the GW only analyses (first prior-posterior pair from the left in each panel), from a sine prior we get a clean, peaked posterior distribution for the viewing angle. In the case of the analysis based on EM data alone (second prior-posterior pair from the left), GW170817 yields a peaked distribution with long tails due to the $\theta_c$--$\theta_v$ degeneracy.  These grow as the distance to the source increases, as shown by the GW170817 simulated event for O4 discussed in the previous section, for which the posterior mildly differs from the prior. Likewise, the value of the jet opening angle $\theta_c$ at 136.5 Mpc has a large scatter and it is recovered as $11\degree^{+13\degree}_{-4\degree}$. The degrading of the measurement of $\theta_v$ and $\theta_c$ with source distance is mainly due to the fact that the rise of the afterglow drops below the sensitivity of all instruments: this part is what places the most stringent constraints on the ratio of the two angles.

The situation improves when the EM data is processed with a prior on $\theta_v$ that is informed by the GW measurement of $\theta_{\rm JN}$.  This is the case of the third prior-posterior pair from the left in both panels, where the prior is 
the posterior of the first case.  This GW-informed prior acts on the $\theta_c$--$\theta_v$ degeneracy: it reduces the tails in the posterior for $\theta_v$ and shifts the median to a lower value.  However, when the source is moved further out (bottom panel), the EM data is not informative and simply allows us to recover the prior.  The results for the EM data with two distinct priors tell us that the prior choice can be crucial, particularly when the afterglow data is not sampled extensively enough throughout the rise, peak, and decay.

Finally, the case of the EM+GW joint fit with a sine prior on the viewing angle gives a
%the sine prior on the viewing angle is enough to extract 
conclusive result on this parameter, in agreement with the EM analysis with the GW-informed prior (as expected). Regarding GW170817, it drives the median to a lower value than the EM-only fit (mostly, as already said above, by ``cutting'' the tails of the $\theta_v$ posterior) and it reduces the width of the distribution by a factor $\sim 1.5$ with respect to the GW-only fit.  
For the GW170817-like event located at 136.5\,Mpc and occurring in O4, these trends are confirmed. 
Both in the EM with GW-informed prior and in the EM+GW analysis, all fits prefer an off-axis observer, but the jet opening angle measurement retains a large scatter, due to the fact that the data points in the rising part of the afterglow are below the sensitivity of the instruments. As mentioned previously, data from this stage of the afterglow strongly constrains the ratio $\theta_v/\theta_c$. For the same reason, for this large distance event, it is hard to derive a conclusive information on the geometry of the jet, as, the values of the jet opening angle and the wing truncation angle $\theta_w$ are consistent at $1\sigma$ with a Gaussian geometry, and above that with a Gaussian geometry with $\theta_w < \theta_c$, which corresponds to a top hat geometry. However, both in the EM-only fit and the EM and GW fits, the jet total width is constrained to be lower than the viewing angle, thus pointing to an off-axis jet.

A summary of the study on the degeneracy between the jet opening angle $\theta_c$ and the viewing angle $\theta_v$ for different distances is represented in Fig.\,\ref{Fig:concl}. Here the 2D posterior distributions for the distant event at 136.5\,Mpc, the intermediate distance event at 70\,Mpc and GW170817 are represented in the first, second and third row respectively (note that GW170817 occurred in O2, while the other events are simulated with the O4 interferometers sensitivities). In the EM fit (first column) the further is the event, the worse is the degeneracy, leaving at 136.5\,Mpc the two angles unconstrained above 1$\sigma$. Including the GW domain (second column) acts only on $\theta_v$, so at large distance $\theta_c$ is still unconstrained above 1$\sigma$ for the reasons we already pointed out. At 70\,Mpc the EM dataset can constrain better these angles, but there is still a strong degeneracy, which is broken by the GW domain. At 41\,Mpc the angles are already very well constrained, thanks to the well sampled light curve, especially at late times in the X-rays, so the GW domain ease the degeneracy, but does not break it.

In O4 we expect to see binary neutron stars mergers up to about 200\,Mpc, in this work we show that, especially at large distances, what is offered by the up to date EM instruments may not be enough to break the degeneracy between the jet opening angle and the viewing angle, so the inclusion of the GW domain is fundamental. In order to observe the rising and declining slope of the afterglow at large distances, instruments with higher sensitivities, such as \textit{Athena} \citep{Piro2021, Piro2022}, are needed, these would allow us to better characterize the properties of the event, extending the EM horizon at higher distances and larger viewing angles.

\section*{Acknowledgements}

We thank Giancarlo Ghirlanda for stimulating discussion.
We acknowledge support by the European Union horizon 2020 programme under the AHEAD2020 project (grant agreement number 871158). This work has been also supported by ASI (Italian Space Agency) through the Contract no. 2019-27-HH.0. Research at Perimeter Institute is supported in part by the Government of Canada through the Department of Innovation, Science and Economic Development and by the Province of Ontario through the Ministry of Colleges and Universities.
This research has made use of data or software obtained from the Gravitational Wave Open Science Center (gw-openscience.org), a service of LIGO Laboratory, the LIGO Scientific Collaboration, the Virgo Collaboration, and KAGRA. LIGO Laboratory and Advanced LIGO are funded by the United States National Science Foundation (NSF) as well as the Science and Technology Facilities Council (STFC) of the United Kingdom, the Max-Planck-Society (MPS), and the State of Niedersachsen/Germany for support of the construction of Advanced LIGO and construction and operation of the GEO600 detector. Additional support for Advanced LIGO was provided by the Australian Research Council. Virgo is funded, through the European Gravitational Observatory (EGO), by the French Centre National de Recherche Scientifique (CNRS), the Italian Istituto Nazionale di Fisica Nucleare (INFN) and the Dutch Nikhef, with contributions by institutions from Belgium, Germany, Greece, Hungary, Ireland, Japan, Monaco, Poland, Portugal, Spain. The construction and operation of KAGRA are funded by Ministry of Education, Culture, Sports, Science and Technology (MEXT), and Japan Society for the Promotion of Science (JSPS), National Research Foundation (NRF) and Ministry of Science and ICT (MSIT) in Korea, Academia Sinica (AS) and the Ministry of Science and Technology (MoST) in Taiwan.

%%%%%%%%%%%%%%%%%%%%%%%%%%%%%%%%%%%%%%%%%%%%%%%%%%
\section*{Data Availability}

The data underlying this article will be shared on reasonable request to the corresponding author.



\bibliographystyle{mnras}
\bibliography{references} 

%%%%%%%%%%%%%%%%% APPENDICES %%%%%%%%%%%%%%%%%%%%%

\appendix
\section{Two datasets, one (or more) shared parameters}
\label{Sec:Appendix}

Suppose that we have two independent datasets $\vec{d}_1$ and $\vec{d}_2$ described by the model \textit{M}. The set of parameters $\vec{\vartheta}$ for \textit{M} is given by parameters exclusive to $\vec{d}_1$, parameters exclusive to $\vec{d}_2$ and a shared parameter between the two datasets: $\vec{\vartheta} = \{ \vec{\vartheta}_1, \vec{\vartheta}_2, \vartheta_0 \}$.
The fit of the two datasets will have a joint likelihood given by the product of the two likelihoods associated to each dataset
\begin{equation}
    \mathcal{L}_{1,2}(\vec{d}_1, \vec{d}_2 |  \vec{\vartheta}, \textit{M}) = \mathcal{L}_{1}(\vec{d}_1 | \vec{\vartheta}, \textit{M} ) \times \mathcal{L}_{2}(\vec{d}_2 | \vec{\vartheta}, \textit{M}),
\end{equation}
here, for notation simplicity, we will drop $\textit{M}$, so
\begin{equation}
    \mathcal{L}_{1,2}(\vec{d}_1, \vec{d}_2| \vec{\vartheta}) = \mathcal{L}_{1}(\vec{d}_1| \vec{\vartheta}) \times \mathcal{L}_{2}(\vec{d}_2| \vec{\vartheta}),
\label{Eq:likelihood_joint}
\end{equation}
The posterior distribution will be proportional to 
\begin{equation}
   p(\vec{\vartheta} |\vec{d}_1, \vec{d}_2) \propto \mathcal{L}_{1,2}(\vec{d}_1, \vec{d}_2| \vec{\vartheta}) \pi(\vec{\vartheta})
\label{eq:posterior_joint}
\end{equation}
where we omit the evidence, which can be treated as a normalization factor. If we marginalize for all the parameters except the shared one, $\vartheta_0$, we get
\begin{equation}
   p(\vartheta_0 |\vec{d}_1, \vec{d}_2) \propto \mathcal{L}_{1,2}(\vec{d}_1, \vec{d}_2| \vartheta_0) \pi(\vartheta_0),
\label{eq:posterior_joint_m}
\end{equation}
where we assume that the prior is separable between the shared parameter, the ones describing only $\vec{d}_1$ and only $\vec{d}_2$: $\pi(\vec{\vartheta}) = \pi(\vartheta_0, \vec{\vartheta_1}, \vec{\vartheta_2}) =  \pi(\vartheta_0)\pi(\vec{\vartheta_1})\pi(\vec{\vartheta_2})$.

Another possibility is to fit one dataset, for example $\vec{d}_1$, and use the resulting marginalized posterior probability of the shared parameter $\vartheta_0$ as prior in the fit of the other dataset, $\vec{d}_2$. This is the case of the EM fit with a GW-informed prior. 
From the fit of the first dataset $\vec{d}_1$ we get a marginalized posterior distribution
\begin{equation}
   p(\vartheta_0 |\vec{d}_1) \propto \mathcal{L}_1(\vec{d}_1| \vartheta_0) \pi(\vartheta_0).
\label{eq:posterior_1}
\end{equation}
where we assume that the prior on the shared parameter is the same as Eq.\,(\ref{eq:posterior_joint_m}). 
For the fit of $\vec{d}_2$ we use Eq.(\ref{eq:posterior_1}) as prior on $\vartheta_0$. The posterior distribution resulting from the fit of $\vec{d}_2$ will be
\begin{equation}
   p(\vartheta_0 |\vec{d}_2) \propto \mathcal{L}_2(\vec{d}_2| \vartheta_0) \pi_2(\vartheta_0),
\label{eq:posterior_2}
\end{equation}
which, since $\pi_2(\vartheta_0) = p(\vartheta_0 |\vec{d}_1)$, becomes
\begin{equation}
   p(\vartheta_0 |\vec{d}_2) \propto \mathcal{L}_2(\vec{d}_2| \vartheta_0)\mathcal{L}_1(\vec{d}_1| \vartheta_0) \pi(\vartheta_0),
\label{eq:posterior_2}
\end{equation}
This posterior is the same as the one found with a simultaneous fit, Eq.(\ref{eq:posterior_joint_m}).
So we can say that when there is one parameter in common, independently on the likelihood function, fitting the two datasets jointly or using the posterior distribution of the shared parameter as prior on that same parameter is equivalent.

In the case of two (or more) shared parameters between two different datasets, the joint fit will give a posterior distribution written in the same way as Eq.(\ref{eq:posterior_joint}), where the likelihood will be given by Eq.(\ref{Eq:likelihood_joint}). Making explicit the shared parameters, which we will call $\vec{\vartheta_s} = \{\vartheta_a, \vartheta_b\}$, and marginalizing for $\vec{\vartheta_1}$ and $\vec{\vartheta_2}$, we get
\begin{equation}
    \mathcal{L}_{1,2}(\vec{d}_1, \vec{d}_2| \vec{\vartheta_s}) = \mathcal{L}_{1}(\vec{d}_1| \vec{\vartheta_s}) \times \mathcal{L}_{2}(\vec{d}_2| \vec{\vartheta_s}),
\label{Eq:likelihood_joint_2p}
\end{equation}
and the posterior will be
\begin{equation}
   p(\vec{\vartheta_s} |\vec{d}_1, \vec{d}_2) \propto \mathcal{L}_{1,2}(\vec{d}_1, \vec{d}_2| \vec{\vartheta_s}) \pi(\vec{\vartheta_s}).
\label{eq:posterior_joint_2p}
\end{equation}
%where the prior distributions are uniform

In the case of the two separate fits, the posterior distribution for the two shared parameters from the fit of $\vec{d}_1$ will be
\begin{equation}
   p(\vec{\vartheta_s} |\vec{d}_1) \propto \mathcal{L}_1(\vec{d}_1| \vec{\vartheta_s}) \pi(\vec{\vartheta_s})
\label{eq:posterior_1_2p}
\end{equation}
The posterior of the fit $\vec{d}_2$ with the prior distribution as in Eq.(\ref{eq:posterior_1_2p}) will be
\begin{equation}
   p(\vec{\vartheta_s} |\vec{d}_2) \propto \mathcal{L}_2(\vec{d}_2| \vec{\vartheta_s}) \pi_2(\vec{\vartheta_s})
\label{eq:posterior_2_2p}
\end{equation}
replacing $\pi_2(\vec{\vartheta_s})$ with its marginalized posterior distributions from the fit of dataset $\vec{d}_1$ we get
\begin{equation}
   p(\vec{\vartheta_s} |\vec{d}_2) \propto \mathcal{L}_2(\vec{d}_2| \vec{\vartheta_s}) \mathcal{L}_1(\vec{d}_1| \vec{\vartheta_s})  \pi(\vec{\vartheta_s}) 
\label{eq:posterior_2_2p}
\end{equation}
which is still identical to the joint posterior in Eq.(\ref{eq:posterior_joint_2p}). However, in this case, we cannot assume the prior is separable between the individual shared parameters:
\begin{equation}
   \pi(\vec{\vartheta_s}) = \pi(\vartheta_a, \vartheta_b) \neq \pi(\vartheta_a) \pi(\vartheta_b),
\label{eq:posterior_2_2p}
\end{equation}
because this requires the parameters to be independent, which is no longer the case after they have been fit to the first dataset.
The new prior 
\begin{equation}
   \pi(\vec{\vartheta_s}) = \int d\vec\vartheta_1 p(\vec{\vartheta_s}, \vec\vartheta_1 | \vec{d}_2)
\label{eq:posterior_2_2p}
\end{equation}
is the posterior from the fit of $\vec{d}_1$ marginalized over only the parameters exclusive to $\vec{d}_1$. This will, in general, be multidimensional and contain all the covariances between the parameters in $\vec{\vartheta_s}$.

% Don't change these lines
\bsp	% typesetting comment
\label{lastpage}
\end{document}